\documentclass[twocolumn]{IEEEtran}
\usepackage{setspace}
\usepackage{amsmath}
\usepackage{amsfonts}
\usepackage{textcomp}
\usepackage{graphicx}
\usepackage{xcolor}
\usepackage{wrapfig}
\usepackage{float}

\renewcommand{\d}[2]{\frac{d#1}{d#2}}
\newcommand{\pd}[2]{\frac{\partial #1}{\partial #2}}

\newcommand{\avg}[1]{\langle #1 \rangle}

\DeclareMathOperator*{\argmax}{argmax}
\begin{document}

\title{Target Detection aided by Quantum Temporal Correlations: Theoretical Analysis and Experimental Validation}
\author{Han Liu, Bhashyam Balaji,~\IEEEmembership{Senior Member,~IEEE,} Amr S.~Helmy,~\IEEEmembership{Fellow,~OSA,}~\IEEEmembership{Senior Member,~IEEE}}
\maketitle
%check consisitence QTC detection protocol
\begin{abstract}
	The detection of objects in the presence of significant background noise is a problem of fundamental interest in sensing. In this work, we theoretically analyze a prototype target detection protocol, the quantum temporal correlation (QTC) detection protocol, which is implemented in this work utilizing spontaneous parametric down-converted photon-pair sources.  The QTC detection protocol only requires time-resolved photon-counting detection, which is phase-insensitive and therefore suitable for optical target detection. As a comparison to the QTC detection protocol, we also consider a classical phase-insensitive target detection protocol based on intensity detection that is practical in the optical regime. We formulated the target detection problem as a total probe photon transmission estimation problem and obtain an analytical expression of the receiver operating characteristic (ROC) curves. We carry out experiments using a semiconductor waveguide source, which we developed and previously reported. The experimental results agree very well with the theoretical prediction. In particular, we find that in a high-level environment noise and loss, the QTC detection protocol can achieve performance comparable to that of the classical protocol (that is practical in the optical regime) but with \(\simeq 57\) times lower detection time in terms of ROC curve metric. The performance of the QTC detection protocol experiment setup could be further improved with a higher transmission of the reference photon and better detector time uncertainty. Furthermore, the probe photons in the QTC detection protocol are completely indistinguishable from the background noise and therefore useful for covert ranging applications. Finally, our technological platform is highly scalable as well as tunable and thus amenable to large scale integration, which is necessary for practical applications.
\end{abstract}

\begin{IEEEkeywords}
	Quantum Lidar, Quantum Radar, Quantum Temporal Correlation, Covert Ranging
\end{IEEEkeywords}

\section*{Notation}

\begin{table}[ht]
\centering % used for centering table
\begin{tabular}{l r } % centered columns (4 columns)
%\hline\hline %inserts double horizontal lines
%$\Delta t_0$& Intrinsic correlation time\\

$\nu$& Photon pair generation rate\\
$\nu_b$& Noise photon detection rate\\
$\Delta t$& Detector time uncertainty\\
$\Delta t_0$& Intrinsic correlation time\\
$\Delta t_{\text{eff}}$& effective time uncertainty\\
$\tau$&Observation time\\
$\eta_r$&Total transmission of reference photons\\
$\eta_p$&Total transmission of probe photons\\
$T_c$& Length of the temporal detection window\\
$t_p$&Probe photon detection time\\
$t_r$&Reference photon detection time\\
$P_p$&Probe photon detection rate\\
$P_r$&Reference photon detection rate\\
$P_c$&Coincidence detection rate\\
$P_d$&Probability of target detection \\
$P_{fa}$& Probability of false alarm\\
$\text{ROC}$&Receiver operator characteristic\\
SPDC&Spontaneous Parametric Downconversion\\
QTI&Quantum Temporal Correlation\\
CTI&Classical Temporal Intensity
\end{tabular}
\label{table:Notation} % is used to refer this table in the text
\end{table}

\section{Introduction}
The fundamental problem in sensing is to detect an object in the presence of interference and noise. Different regions of the electromagnetic spectrum are exploited for such sensing applications; those include, radar in the radiofrequency regime and LIDAR in the optical regime. Current radars/LIDARs are based on signals that can be described by classical properties of electromagnetic radiation. These properties pose some limitations on the sensing performance including thermal and shot noise.
A novel fashion that can be devised to overcome such performance limits is to utilize non-classical sources of electromagnetic radiation.
This is possible because quantum physics allows for other types of states of electromagnetic radiation that exploit two key properties in quantum mechanics, namely \emph{the quantization of the electromagnetic field}\cite{greiner2013field} (where the quantized excitations are referred to as \emph{photons}) and  \emph{quantum entanglement and correlations}\cite{einstein1935can}\footnote{Correlation between two particles means that a certain degree of freedom of the two particles always takes correlated values.
Quantum entanglement between two particles means that the two particles are correlated in different degrees of freedom, in such a way that it is not possible to describe each of the two particles separately.}.
It has been experimentally demonstrated that quantum entanglement \cite{lloyd2008enhanced} and correlations\cite{lopaeva:2013,BBDE2018,chang2019quantum} that exist within entangled photon pairs could be utilized to enhance the accuracy of target detection in lossy and noisy environments.
The basic idea of these protocols can be described as follows:
A pair of nonclassical photons in either the radio or optical frequency regimes are used as the source, which consists of the \emph{probe} and the \emph{reference} photons (also referred to in the literature as \emph{signal} and \emph{idler} photons, respectively).
The probe photons are sent towards the target and the back-reflected ones are collected for detection.
The reference photons are stored or immediately detected locally.
Through analyzing the quantum entanglement, or correlation, between the probe and the reference photon with appropriate measurement scheme, it is possible to reduce the environmental noise effect on the target detection performance.
This is because environmental noise photons are not correlated or entangled with the reference photons.   \\

An example of quantum-enhanced target detection protocols is \emph{quantum illumination}, which can effectively distinguish the presence and absence of the target object through analyzing the entanglement between the reference and the probe light. It has been shown that the sensitivity of the quantum illumination protocol can surpass the classical limit that is achieved by phase-sensitive homodyne detection \cite{Zhang:2015}.
However, the implementation of the quantum illumination protocol requires a significant level of system complexity, including phase-sensitive joint detection\footnote{a joint measurement requires bringing the two photons to the same physical location at the same time.}. For example, the optical phases between the probe and reference light have to be stabilized down to the sub-wavelength level \cite{Zhang:2015}, which is not practical if the target distance is unknown or fluctuating. As such, formidable challenges lie in implementing the quantum illumination protocol to benefit practical target detection applications. An alternative approach to enhancing the performance of target detection, while mitigating the complexity of quantum illumination, is to only use correlations that exist in the non-classical states of light. The correlation enhanced protocols only require independent, phase-insensitive measurements to analyze the correlation between the probe and reference light. The performance of correlation enhanced phase-insensitive target detection protocols may be inferior to the optimal classical target detection protocol based on phase-sensitive detection, it is nevertheless useful for practical phase-insensitive target detection, such as radar and LIDAR. The key aspect of correlation enhanced target detection protocols is the type of correlation that is utilized. One type of correlation that exists in entangled photon pairs is the \emph{temporal correlation}, which means the detection time of the probe and reference photons are completely random but always `simultaneous'.
In other words, the probe and the reference photons are always detected with the detection time difference smaller than the \emph{intrinsic correlation time} \(\Delta t_0\)(assuming zero detector time uncertainty), which could be as short as tens of femtoseconds\cite{Abolghasem:2009}.
To measure the temporal correlation, it suffices to conduct separate phase-insensitive time-resolving photon counting detection for the reference and probe photons.
Recent years have seen rapid progress in single photon detection technology with ultra short detection time uncertainty down to picosecond level\cite{caloz2018high}, which has made possible the utilization of the strong temporal correlation of non-classical photon pairs.

In this paper, we discuss an approach to target detection (the QTC detection protocol) that utilizes the temporal correlation of non-classical photon pairs that are generated in a monolithic semiconductor waveguide.
The rest of this paper is organized as follows:
In Section II the basic formalism of the QTC detection protocol is explained and is compared to a similar protocol in the radio frequency domain. The QTC detection protocol is also compared to a classical intensity detection based (CTI) protocol that is commonly used in the optical domain. In Section III a simple theoretical model for the QTC detection protocol is developed. In particular, the parameter estimation theory is applied to construct the optimal detector function for the QTC detection protocol, based on which the theoretical prediction of the ROC curve is obtained. In Section IV the experiment setup is described. Section V shows the experimentally measured time series of the detector functions of both the QTC and CTI detection protocol, from which the ROC curve is calculated. The experiment result is compared to the theoretical prediction which is obtained in section IV. In section VI the limiting factors of the QTC detection protocol performance are discussed and the QTC detection protocol is compared to the CTI detection protocol with pulsed probe light and other correlation enhanced protocols. The conclusion is discussed in Section VII.

\section{Background and Overview}
The goal of this paper is to develop and analyze a phase-insensitive target detection protocol (the QTC detection protocol) that utilizes the strong temporal correlations of non-classical photon pairs.
This enhanced protocol will be compared to a classical and practical optical target detection protocol (the CTI detection protocol) based on time-resolved intensity (photon-counting) detection.
Both the enhanced and the baseline classical target detection experiments are conducted in a lossy and noisy environment, where the environment noise is assumed to be overlapping with the probe signal in both temporal and spectral domain (i.e., background photons are in the same frequency range as the signal and could also arrive at the same time as a signal photon).

\begin{figure}[ht]
\centering
\includegraphics[width=0.8\columnwidth]{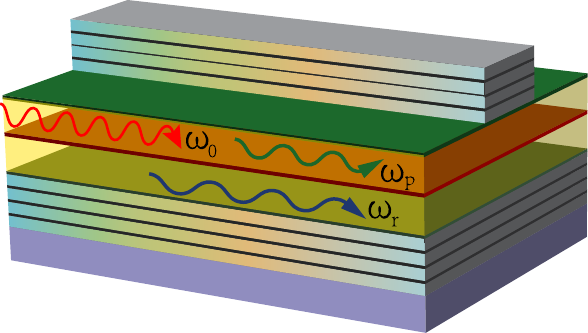}
\includegraphics[width=0.8\columnwidth]{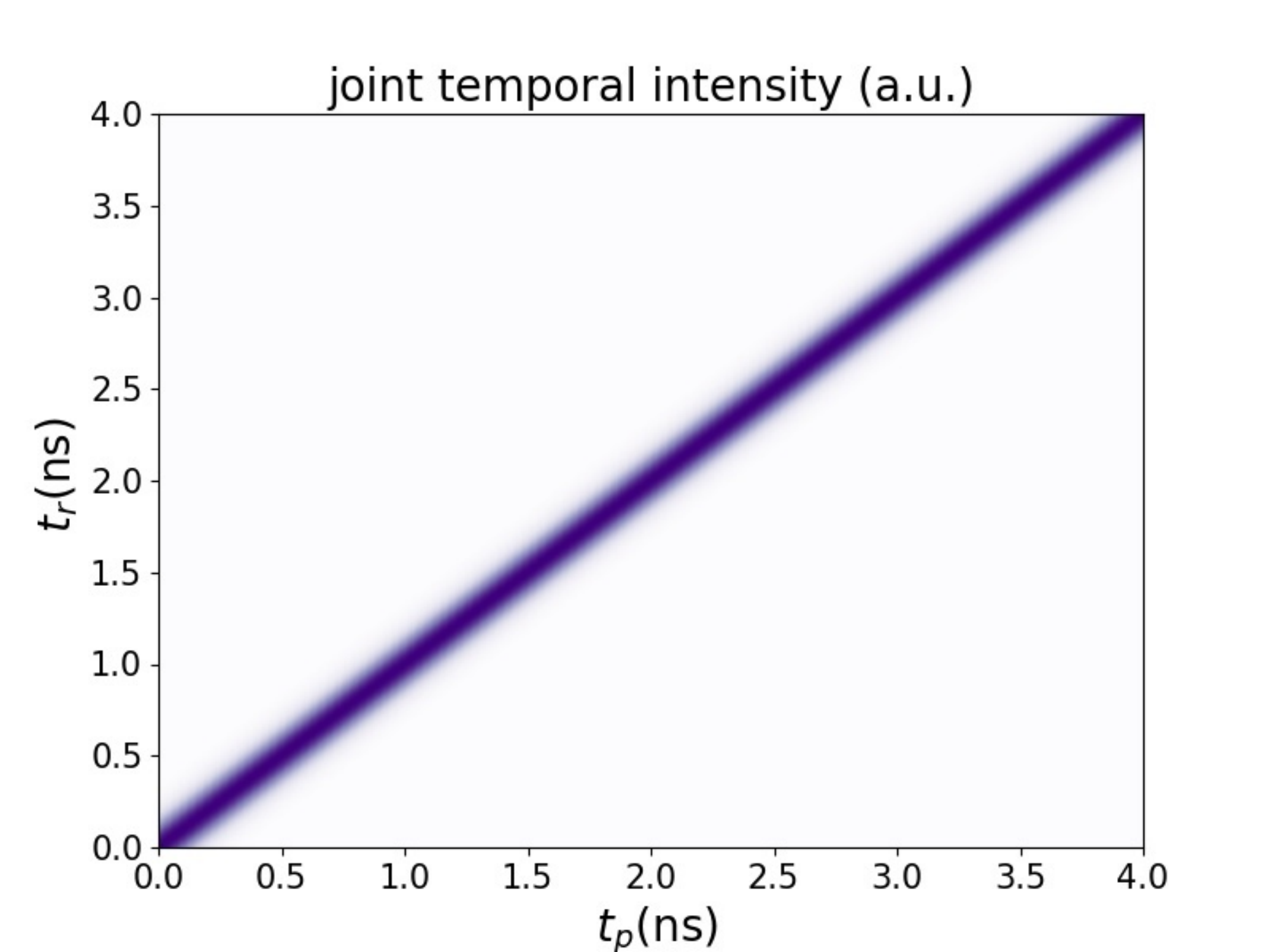}
\includegraphics[width=0.8\columnwidth]{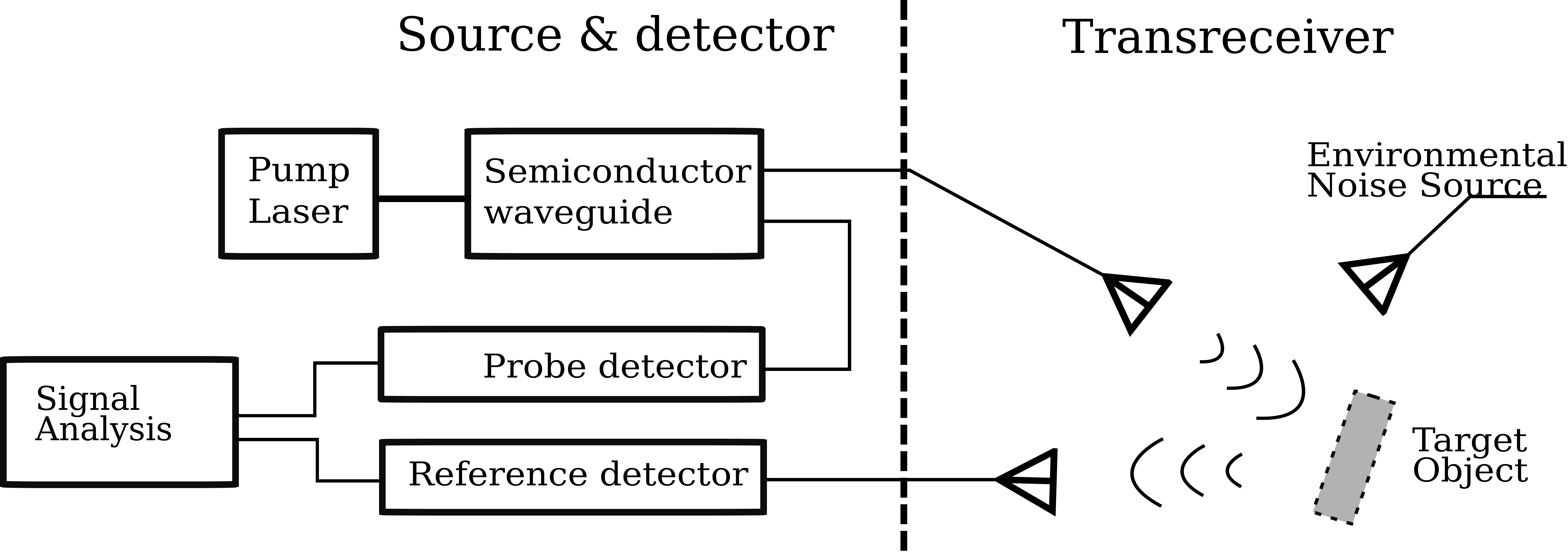}
\caption{ Top: the schematic of the SPDC process in the semiconductor waveguide. The red wave \(\lambda_0\) is the pump light at 783nm, the green \(\lambda_p\) and blue \(\lambda_r\) wave is the down-converted  probe and reference photon that is in horizontal (vertical) polarization. Middle: joint probability distribution of photon detection time.
\(t_p\): detection time of the probe photon; \(t_r\) detection time of the reference photon.
The standard deviation of the detection time difference is \(\Delta (t_p-t_s) = 100ps\) for this plot. Bottom: Block diagram of the experiment setup.
}\label{JTI}

\end{figure}

\subsection{Temporal Correlations}
In the QTC detection protocol, the temporal correlation between the probe and the reference photon is utilized. Temporally correlated photons could be generated through continuous-wave (CW) pumped spontaneous parametric down-conversion (SPDC)\cite{klyshko1970parametric,zhukovsky2013analytical}.
In the SPDC process, a pump photon with a short wavelength (\(\simeq 783\) nm) is annihilated in the nonlinear medium and subsequently generates the probe and reference photon pair with longer central wavelength  (\(\simeq 1566\) nm).
While the bandwidth of the CW pump light is narrow, the generated probe and reference photons could be broadband, depending on the structure of the semiconductor waveguide \cite{Abolghasem:2009}. This is because the annihilation of the pump photon could generate a photon pair with different possible frequency combinations, as long as the sum of the frequencies of the photon pairs equals the frequency of the pump.
The probe and reference photons are in different polarization states (i.e., Type II SPDC) and are separated by a polarization beamsplitter.
Note that the photons in the SPDC photon pair are entangled in polarization before the polarization splitting. However, in this paper, we are only considering temporal correlations between the probe and the reference photon after the polarization splitting.
Since the pump light is CW, the probe and reference photons are generated at completely random times.
The temporal correlation of the probe and reference photon could be resolved via time-resolving photon detection:
the detection time difference of the probe and reference photon is always smaller than the \textbf{intrinsic correlation time}, \(\Delta t_0\), of the photon pair, if both photons have traveled equal optical path length and the detectors have the perfect temporal resolution.
This corresponds to the heuristic idea that the probe and the reference photons are always generated at the `same time'.
The typical joint probability distribution of probe and reference photon detection time is shown in Fig. \ref{JTI}.\\

\subsection{Experimental details of the QTC detection protocol}
The block diagram of the QTC detection experiment setup is shown in Fig. \ref{JTI}.
After the generation of the probe and the reference photon in the nonlinear-waveguide, the reference photon is immediately detected on a single photon detector (the reference detector) and the probe photon is sent toward the target object.
The probe photons which are back-reflected from the target object are collected and detected on another single-photon detector (the probe detector).
Regardless of the presence or absence of the target object, the strong environmental noise power is always detected on the probe detector.
The strong background noise is assumed to be completely overlapping with the probe photons in terms of both spectral and temporal domain distribution, i.e. CW and broadband.
Therefore for the cases discussed here, it is not possible to reduce the noise power through temporal or spectral filtering.
The presence and absence of the target object could be determined by analyzing the time-resolved photon detection statistics on both detectors.
Since the background noise photons are uncorrelated with the reference photon, the effect of the background noise on the target detection accuracy could be reduced by analyzing the temporal correlation information that is extracted from the photon detection statistics.
The CTI detection protocol is a simple intensity-based target detection protocol with experiment setup identical to that of the QTC detection protocol, except that the reference photons are not detected (nor is the information used in the processing). In other words, the CTI protocol simply detect the intensity of the reflected probe light to determine the presence of the target object, without using any correlation information. It is true that the CTI protocol is not the optimal classical target detection protocol and it can be much improved using many other classical techniques such as coherent detection and matched filtering. However, since intensity detection has already be widely adopted for optical target detection, the CTI protocol can still be chosen as a baseline for comparison for the QTC detection protocol, at least in the optical regime.

\subsection{Experimental Parameters}
There are some important characteristic constants of the QTC detection protocol and the CTI detection protocol that are closely related to their performance.  These constants will be used in both the theoretical analysis and the experimental data analysis.
\begin{itemize}
\item The \textbf{source pair rate}, \(\nu\), is defined as the number of SPDC photon pairs generated inside the waveguide per second.\\

\item The measured temporal correlation of the probe and the reference photon depends on two factors:
\begin{itemize}
\item \textbf{Detector time uncertainty (\(\Delta t\)}): assuming the reference and probe single-photon detector have identical performance, \(\Delta t\) is defined as twice the maximal difference between the recorded photon detection time and the actual photon detection time.\\

\item\textbf{Intrinsic Temporal Correlation (\(\Delta t_0\)}): it is defined as the maximal time difference between the detection of the probe and the reference photons, assuming zero detector time uncertainty.\\
\end{itemize}

The \textbf{effective time uncertainty}, \(\Delta t_{eff}\), of photon detection is defined as the sum of detector time uncertainty \(\Delta t\) and intrinsic correlation time \(\Delta t_0\):
\begin{align}
\Delta t_{eff}=\Delta t+\Delta t_0.
\end{align}
\item The \textbf{total reference photon transmission efficiency} \(\eta_r\) is defined as the ratio between the number of photon detection rate on the reference detector and the source pair rate \(\nu\), i.e.,
\begin{align}
\eta_r=\frac{\text{Reference photon detection rate}}{\text{Photon source pair rate}}.
\end{align}
The total reference photon transmission efficiency \(\eta_r\) includes the effect of both the optical loss of the reference photons and the reference detector efficiency.\\

\item The \textbf{total probe photon transmission efficiency }, \(\eta_p\), is similarly defined as (assuming no background noise)
\begin{align}
    \eta_p=\frac{\text{Probe photon detection rate}}{\text{Photon source pair rate}}.
\end{align}
Note that $\eta_p$ is affected by the detection efficiency and the coupling losses as well. In addition, \(\eta_p\) is affected by the distance of the target object, the reflectivity of the object and the collection efficiency of back-reflected probe photons.\\

\item The power of background noise \(\nu_b\) is characterized by the number of photon detection events per second that is due to the noise source (i.e. with probe beam blocked).\\

\item Another important parameter is the \textbf{coincidence detection window}, \(T_c\), relating to the definition of \textbf{coincidence detections}, which will be made clear in the following section.\\
\end{itemize}

\section{Theoretical analysis}
\subsection{Temporally correlated photon pairs}
In order to exploit the temporal correlation, it suffices to model the correlated photon pair states with the joint probability density \(P(t_p,t_r)\) of generating a probe photon at time \(t_p\) and a reference photon at time \(t_r\).
The average number of the photon pairs generated per second \(\nu\) equals the average number of reference photons generated per second, i.e.,
\begin{gather}
\nu = \int_{-\infty}^{+\infty}dt_pP(t_p,t_r)
\end{gather}
For CW pumped SPDC process, \(P(t_p,t_r)\) only depends on the time difference \(t_p-t_r\). Therefore \(\nu\) does not depend on \(t_r\)(despite the explicit appearance of \(t_r\) in the above expression).
The intrinsic correlation time \(\Delta t_0\), which is defined as the maximal detection time difference \(t_p-t_r\), could be approximated as 3 times the standard deviation of the time difference \(t_p-t_r\):
\begin{gather}
\Delta t_0 = 3\times\sqrt{\frac{1}{\nu }\int_{-\infty}^{+\infty}dt_r P(t_p,t_r)(t_p-t_r)^2}
\end{gather}
The intrinsic correlation time \(\Delta t_0\) is determined by the spectral property of the SPDC photon pair and could be tuned from around 20 fs to around 10 ps through different waveguide design\cite{Abolghasem:2009}.
The joint probability density distribution of the photon detection time is a convolution between the joint probability density of the photon generation time \(P(t_p,t_r)\) and the detector response function (parametrized by the detector time uncertainty \(\Delta t\)), aside from a constant time shift of \(t_p\) or \(t_r\) that is due to the unbalanced optical path lengths of the probe and the reference photon.
Such unbalance could be compensated for through post data processing and is irrelevant for target detection.
Therefore in the rest of the paper, we shall assume that the optical path length difference between the probe photon and the reference photon is zero.
However, it should be noted that this time shift contains the target distance information and is useful for ranging-detection. The ability of ranging of the QTC detection protocol is discussed in section VI.C.

\subsection{Photon detection statistics of the QTC detection protocol}
In the QTC detection protocol experiment, the reference photons are detected on the reference detector and the back-reflected probe photon along with background noise photons are detected on the probe detector.
The time of each photon detection event is recorded.
The absolute time of each photon detection event is not useful since the target detection system is time-invariant: the nonclassical photon pair source is CW pumped and the target object is stationary.
For a QTC detection experiment that last for a fixed period of time \(\tau\), there are only three useful quantities that can be extracted from the photon detection statistics:

\begin{itemize}
\item Number of single channel photon detection events \(\mathbf{N}_p\) on the probe detector\\

\item Number of single channel photon detection events \(\mathbf{N}_r\) on the reference detector\\

\item Number of coincidence detection events \(\mathbf{N}_c\) that depends on the time difference between the photon detection events on different detectors.
\end{itemize}
A \emph{coincidence detection} event is defined as detecting two photons on the different detectors `almost simultaneously', with the detection time \(t_p\) and \(t_r\) satisfying:
\begin{gather}
 t_p-t_r \in \left[-\frac{T_c}{2}, -\frac{T_c}{2}\right]
\end{gather}
where \(T_c\) is the length of the coincidence window.
The single channel detection events on the probe or reference detector are defined as the photon detection events that do not contribute to the coincidence detection events.\\

It can be shown (see the Appendix) the \(\mathbf{N}_p\), \(\mathbf{N}_c\) and \(\mathbf{N}_r\) each satisfy independent Poisson distribution:
\begin{gather}
P(\mathbf{N}_p,\mathbf{N}_r,\mathbf{N}_c) =\quad f(\mathbf{N}_p,P_p\tau)f(\mathbf{N}_r,P_p\tau)f(\mathbf{N}_c,P_p\tau)\label{JOINTPDF}
\end{gather}
where
\begin{gather}
f(k,\lambda) = e^{-\lambda}\frac{\lambda^k}{k!}\label{poisson}
\end{gather}
is the Poisson distribution function and the average number of photon detections event \(\mathbf{N}_p, \mathbf{N}_r,\mathbf{N}_c\) per unit time \(P_p\), \(P_r\) and \(P_c\) are given by:
\begin{align}\label{PcEXPR}
P_p &= \nu\eta_p+\nu_b - P_c,\\ \nonumber %\label{Pp_expr},\\ \nonumber
P_r &= \nu\eta_r- P_c,\\ \nonumber
P_c &= \nu\eta_p\eta_r+\nu\eta_r\nu_bT_c. \hspace{12pt} (T_c\ge2\Delta t_{eff})
\end{align}

The above expressions for the photon detection event rates are rigorously derived through quantum mechanical analysis\cite{liu2019enhancing}. A simplified derivation that is only based on the joint temporal probability density \(P(t_s,t_i)\) could be found in the Appendix.
The condition for \eqref{PcEXPR} to hold is that the temporal detection window \(T_c\) is larger than twice the effective time uncertainty \(t_{eff}\).
This is to ensure that probe-reference photon pairs that are detected will always contribute to a coincidence detection event (see the Appendix).
Equation \eqref{PcEXPR} could be used to predict the experimentally measured photon detection rates as well as to calculate experiment parameters \(\nu,\eta_p,\eta_r\) from experimental photon detection statistics.

\subsection{Transmission estimation and detector functions}
For the QTC detection protocol, it is in principle possible to infer the presence or absence of the target object directly from the experimental photon detection statistics \eqref{JOINTPDF}.
However, the fact that \eqref{JOINTPDF} is a multidimensional probability distribution prevents the direct application of standard target detection analysis such as ROC analysis, which requires a single-valued target detection signal as the input. Therefore, in order to quantify the performance of the QTC detection protocol, one must first construct a suitable single-valued detector function, which is a real function of the experimentally measured photon counting statistics \(\mathbf{N}_p,\mathbf{N}_r\) and \(\mathbf{N}_c\).\\

Our approach to constructing the optimal detector function for the QTC detection protocol is to formulate the target detection problem as a probe photon transmission \(\eta_p\) estimation problem\cite{liu2019enhancing}. An unbiased estimator \(\mathbf{S}\) of \(\eta_p\) is a function of experimental photon detection statistics \(\mathbf{N}_p,\mathbf{N}_r,\mathbf{N}_c\) and have its mean value equal the probe transmission \(\eta_p\), which could be formally expressed as:
\begin{align}
\text{\textbf{target present  }:}& \mathbf{S}=\eta_p+\mathbf{n}(\eta_p)\\ \nonumber
 \text{\textbf{target absent  }:}& \mathbf{S}=\mathbf{n}(0)
\end{align}
where $\mathbf{n}(\eta_p)$ and $\mathbf{n}(0)$ are corresponding zero-mean noise variance when the object is present and absent (note that \(\eta_p=0\) when the target object is absent). The unbiased estimator \(\mathbf{S}\) could be used as the detector function. This is because the target object is present if and only if the mean value of \(\mathbf{S}\) is positive. The performance of the detector function \(\mathbf{S}\) is given by the estimation variance, which characterizes the minimal probe photon transmission efficiency \(\eta_p\) that could be distinguished from zero transmission. The performance of the detector function is also affected by the value of total probe transmission \(\eta_p\) itself, which is the same for all detector functions.\\

In this article, we will consider three different types of estimators (detector functions):
\begin{align}
\mathbf{S}_\text{QTC1}& = \argmax\limits_{\eta_p} P(\mathbf{N}_p,\mathbf{N}_r,\mathbf{N}_c)\label{DETS} \\ \nonumber
\mathbf{S}_\text{QTC2} &= \argmax\limits_{\eta_p} f(\mathbf{N}_c,P_c\tau)\\ \nonumber
\mathbf{S}_\text{CTI} &= \argmax\limits_{\eta_p} f(\mathbf{N}_p,P_p\tau) \hspace{1cm}(\eta_r=0)
\end{align}
where \(\mathbf{N}_p,\mathbf{N}_r,\mathbf{N}_c\) are the experimentaly measured photon detection statistics and \(f(\mathbf{N}_p,P_p\tau),f(\mathbf{N}_c,P_c\tau),P(\mathbf{N}_p,\mathbf{N}_r,\mathbf{N}_c)\) are the corresponding probability distribution as given in \eqref{JOINTPDF} and \eqref{poisson}.
Note that $\mathbf{S}_\text{CTI}$ does not use the information about the reference photon (\(\eta_r=0\)), and therefore corresponds to the CTI detection protocol.\\

In the limit of long detection time \(\tau\), the minimal estimation variance could be achieved by QTC1, which is the maximal likelihood estimation (MLE) of \(\eta_p\) from the experimental photon counting statistics \(\mathbf{N}_p,\mathbf{N}_r\) and \(\mathbf{N}_c\)\cite{ly2017tutorial}:
\begin{align}
\avg{\Delta^2\mathbf{S}_\text{QTC1}} =\frac{1}{I_\text{QTC1}},\label{CRAMER_RAO}
\end{align}
where
\begin{align}\label{FI}
I_\text{QTC1} &= \sum\limits_{N_p,N_r,N_c=0}^{+\infty} P(\mathbf{N}_p,\mathbf{N}_r,\mathbf{N}_c)\\ \nonumber
&\quad \times\left[\pd{}{\eta_p}\text{log}P(\mathbf{N}_p,\mathbf{N}_r,\mathbf{N}_c)\right]^2.%\label{FI}
\end{align}
It therefore follows that (see the Appendix for the derivation)
\begin{align}\label{FIEXPR}
I_\text{QTC1}= \left[\frac{\eta_r^2\nu^2}{P_c}+\frac{(1-\eta_r)^2\nu^2}{P_p}+\frac{\eta_r^2\nu^2}{P_r}\right]\tau.
\end{align}
The three terms form left to right in the above expression represent the contribution to the total Fisher information from coincidence detection events \(\mathbf{N_c}\), single channel detection events on the probe and reference detector \(\mathbf{N}_p\),\(\mathbf{N}_{r}\) , as indicated by their respective denominator \(P_c,P_p\) and \(P_r\). When the target detection environment is noisy (\(\nu_b\gg\nu\eta_p\)), the contribution of coincidence detection \(\mathbf{N}_c\) dominates in expression \eqref{FIEXPR}, since the denominator in the first term \(P_c\) is much lower than the denominator in the second term \(P_p\).
This is because the environmental noise will directly increase the single-channel detection rate \(P_p\) but insignificantly affect the coincidence detection rate \(P_c\).
From \eqref{FIEXPR} it can be seen that the effect of the environmental noise \(\nu_b\) to the coincidence detection rate \(P_c\) decreases as the coincidence window size \(T_c\) decreases.  The minimal coincidence window \(T_c\) is limited by twice the effective time uncertainty \(\Delta t_{eff}\).
This implies that the performance of the QTC1 detector function depends on the temporal correlation of non-classical photon pairs as well as the detector time uncertainty.
In most cases, the detector time uncertainty \(\Delta t\) is orders of magnitude larger than the intrinsic correlation time \(\Delta t_0\).
Therefore the effective time uncertainty \(\Delta t_{eff}\) is effectively just the detector time uncertainty, i.e.,  \(\Delta t_{eff}\approx \Delta t\).
In the limit of zero coincidence window (zero effective time uncertainty \(T_c=2\Delta t_{eff} = 0\)), the contribution to the total Fisher information from the coincidence detection events \(\mathbf{N}_c\) is immune to environment noise.\\

The detector function  \(\mathbf{S}_\text{QTC2}\) could be considered as the MLE of \(\eta_p\) when only the marginal probability distribution of coincidence detections \(N_c\) is considered.
\begin{align}\label{FI_EXPR_NC0}
    \mathbf{S}_\text{QTC2} &= \argmax\limits_{\eta_p} f(\mathbf{N}_c,\tau P_c),\\ \nonumber
           &= \argmax\limits_{\eta_p}\{\exp(-(\eta_p\eta_r\nu+\nu\eta_r\nu_bT_c))\\
&\times\frac{(\eta_p\eta_r\nu+\nu\nu_b\eta_rT_c)^{N_c}}{N_c!}\}\nonumber\\
           &=\frac{\mathbf{N}_c-\nu_b\nu\eta_rT_c}{\nu\eta_r},\label{MLE_NC}.%\\ \nonumber
\end{align}
The corresponding Fisher information of \(\mathbf{S}_\text{QTC2}\) is given by \(I_\text{QTC2}\)(see the Appendix for the detailed derivation), which equals the first term of \(I_\text{QTC1}\) in \eqref{FIEXPR}:
\begin{align} \label{FI_EXPR_NC}
I_\text{QTC2} &= \frac{\eta_r^2\nu^2}{P_c}\tau.
\end{align}\\
The CTI detector function \(\mathbf{S}_{CTI}\) is the MLE of the probe transmission efficiency \(\eta_p\) from the marginal distribution of probe channel counts \(N_p\) (when \(\eta_r = 0\)), and the corresponding detector function \(S_\text{CTI}\) and Fisher information (\(I_\text{CTI}\)) are given by
\begin{align}
    \mathbf{S}_\text{CTI}&=\argmax\limits_{\eta_p}f(\mathbf{N}_p,\eta_p\nu+\nu_b),\\
           &=\frac{\mathbf{N}_p-\nu_b}{\nu},\\
    I_\text{CTI} &= \frac{\nu^2}{P_p}\tau.\label{FI_EXPR_CD}
\end{align}
The derivation of \(\mathbf{S}_\text{CTI}\) and \(I_\text{CTI}\) is similar to the derivation of \(\mathbf{S}_\text{QTC2}\) and \(I_\text{QTC2}\).
The detector function \(\mathbf{S}_\text{QTC2}\) and \(\mathbf{S}_\text{CTI}\) are equivalent to more straight forward definitions \(\mathbf{S'}_\text{QTC2}=\mathbf{N}_c\) and \(\mathbf{S'}_\text{CTI} = \mathbf{N}_p\) in terms of ROC analysis as will be shown later, since they are merely re-parametrizations of each other. The detector function \(\mathbf{S'}_\text{QTC2}\) has already been used in a previous study \cite{ BBDE2018}.\\

\subsection{Gaussian approximation and the ROC curve}
The central limit theorem suggests that if the measurement time \(\tau\) of a target detection experiment is sufficiently long, the detector functions \(\mathbf{S}_\text{QTC1},\mathbf{S}_\text{QTC2}\) and \(\mathbf{S}_\text{CTI}\) approximately obeys normal distribution. Therefore, under the large \(\tau\) limit, the noise variance \(\mathbf{n}(\eta_p)\) of the detector functions \(\mathbf{S}_\text{QTC1},\mathbf{S}_\text{QTC2}\) and \(\mathbf{S}_\text{CTI}\) could be approximated as zero-mean Gaussian noise random variables as follows:
\begin{gather}
\mathbf{n}_\text{QTC1}(\eta_p)\sim\text{normal}(0,1/I_\text{QTC1})\label{GAPPROX}\\\nonumber
\mathbf{n}_\text{QTC2}(\eta_p)\sim\text{normal}(0,1/I_\text{QTC2})\\\nonumber
\mathbf{n}_\text{CTI}(\eta_p)\sim\text{normal}(0,1/I_\text{CTI})
\end{gather}
It is important to remember that the Fisher information (\(I_\text{QTC1}\), \(I_\text{QTC2}\) and \(I_\text{CTI}\)) are functions of \(\eta_p\). In particular, \(\eta_p=0\) when the target object is absent. For a fixed target detection time \(\tau\), the validity of this Gaussian signal approximation should be verified with the experimental photon detection statistics.\\

For target detection, the presence and the absence of the target could be determined by comparing the detected signal (the value of the detector function for a particular experiment) to a signal intensity threshold \(V\): when the detected signal is stronger than the threshold the presence of the target object could be asserted and vice versa.
However, even if the object is absent (\(\eta_p=0\)), there could be non-vanishing probability (false alarm possibility \(P_{fa}\)) of the detected signal being higher than the detection threshold. Similarly, if the target object is present (\(\eta_p>0\)), the probability (detection probability \(P_d\)) of the detector function being higher than the threshold could be lower than unity too.
The false alarm rate (\(P_{fa}\)) and detection rate (\(P_{d}\))both depend on the threshold \(V\).
By varying the threshold \(V\), the relationship between \(P_{fa}\) and \(P_{d}\) could be calculated, which is defined as the ROC curve.
The Gaussian signal approximation could be used to obtain an analytic expression of the ROC curve.
Consider any one of the detector functions in \eqref{DETS}.
Denote the Fisher information \(I_\text{QTC1}\) (or \(I_\text{QTC2}\), \(I_\text{CTI}\)) by \(I(\eta_p)\) as the function of total probe transmission \(\eta_p\) (note that \(I(0)\) correspond to the absence of the target).
Then the expression of the ROC curve is given by\cite{marzban2004roc}:
\begin{gather}
P_d = \Phi\left(\eta_p\sqrt{I(\eta_p)}+\sqrt{\frac{I(\eta_p)}{I(0)}}\Phi^{-1}(P_{fa})\right)
\end{gather}
where \(P_d\) is the detection rate and \(P_{fa}\) is the false alarm rate and \(\Phi\) is the cumulative Gaussian distribution function:
\begin{gather}
\Phi(x) = \frac{1}{\sqrt{2\pi}}\int_{-\infty}^{x} \exp(-z^2/2)dz\label{ROC_CURVE}
\end{gather}
The ROC curve is independent of the re-parametrization of detector functions\cite{marzban2004roc}: if a detector function could be expressed in terms of another detector function and constants, then their corresponding ROC curve will be identical.

\begin{table}[!h]
\centering
\caption{Table of some experiment parameters}\label{PARAMS}
\begin{tabular}{|c|c|}
\hline
Transmit Power & 0.41 pW\\
Range to Target& 8 in\\
Source  (\(\nu\))& 3.19 MHz\\
 Background (\(\nu_b\)) &0.85 MHz\\
 Transmission (\(\eta_p\)) &0.24\% \\%2.4\textperthousand\\
 Transmission (\(\eta_r\))&17\% \\
 Window (\(T_c\) )&486 ps\\
Detector time uncertainty ( \(\Delta t\)) &243 ps\footnote{include various factors such as the temporal jittering of the voltage adapters.} \\
 Intrinsic correlation time (\(\Delta t_0\)) &0.02 ps\\
Effective time uncertainty (\(\Delta t_{eff}\)) &\(\simeq\)243 ps\\
 Observation time (\(\tau\)) &0.01 s\\ \hline
\end{tabular}
\end{table}

\section{Experimental setup}
\begin{figure*}[ht]
\centering
\includegraphics[width=1.6\columnwidth]{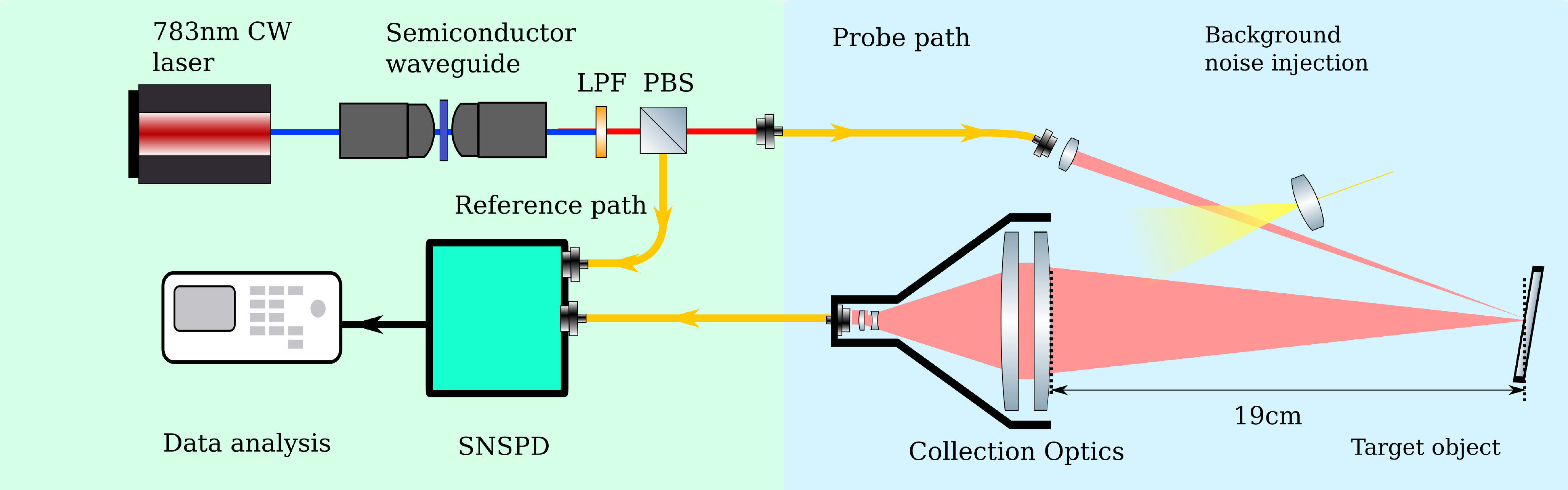}
\caption{The schematic of the experimental setup is divided into two parts: the source and detection part and the transceiver part. The source and detection part (green background) includes the photon pair source and detectors. Pump laser: Ti-Sapphire CW laser at 783nm. PBS: polarization beam splitter. LPF: long pass (\(>\)1200 nm) filter to separate the SPDC photon pairs from the pump laser. SNSPD: dual-channel superconducting single-photon detector (top channel: the reference detector, bottom channel: the probe detector). Transceiver part (blue background): probing of the target object and collection of the back-reflected photons. Target object: a piece of aluminum foil with diffused reflection. The source and detector part and the transceiver part are built on a separate table and are connected by single-mode fibers (yellow line with arrow).}\label{SETUP}
\end{figure*}
\begin{figure}[h]
\centering
\includegraphics[width=0.8\columnwidth]{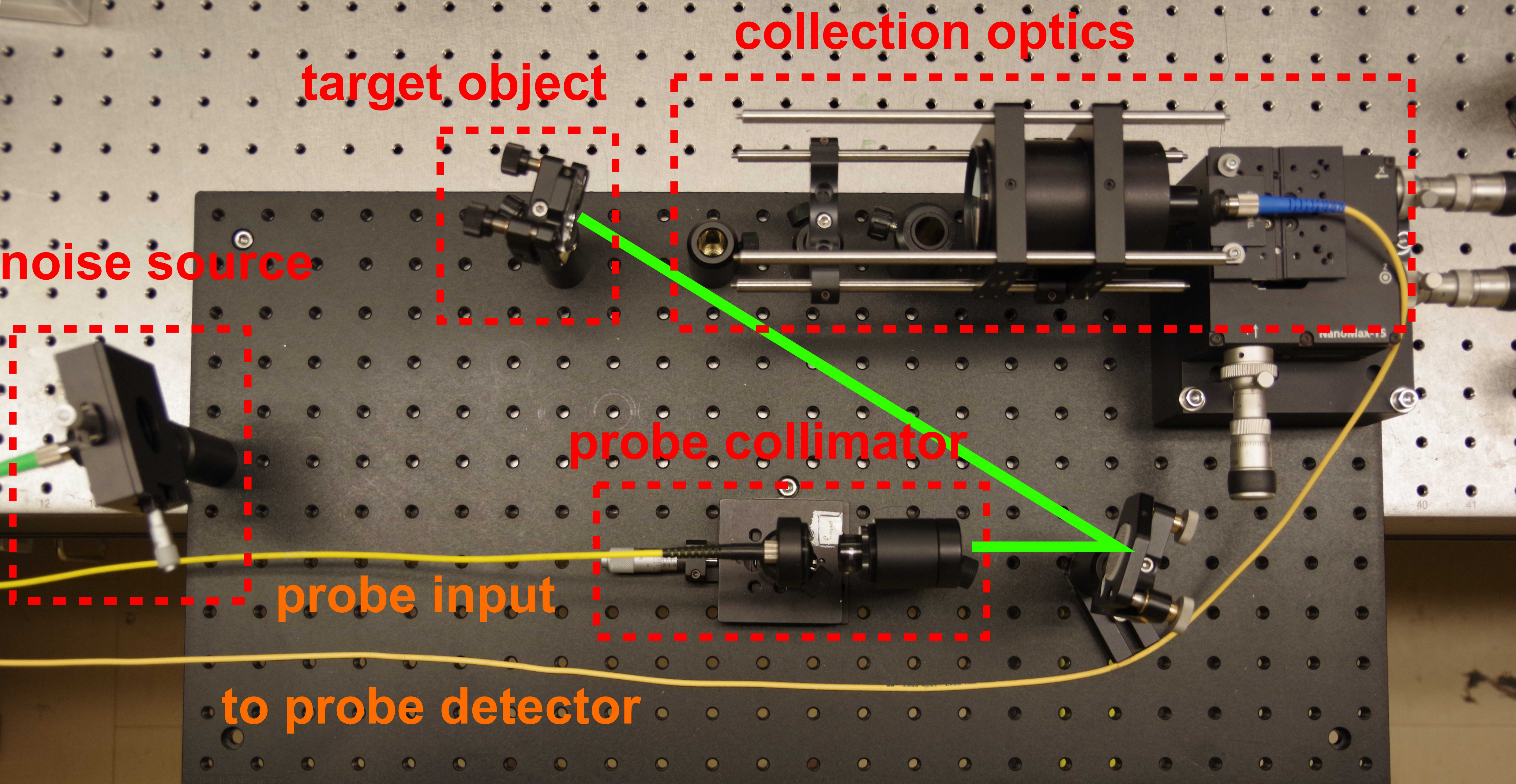}
\caption{Photograph of the transceiver part of the experiment setup. Different optical elements are marked with red boxes. The green line marks the optical path of the probe photon before it hits on the target object. The reflective mirror on the bottom-right corner is purely for optical alignment purposes.\label{PIC}}
\end{figure}
\subsection{Layout}
As shown in the block diagram in Fig. \ref{JTI} and the experiment setup Fig. \ref{SETUP}, the QTC detection protocol setup consists of two parts: the \textbf{source and  detector} part and the \textbf{transceiver part}.
The source and detector part consists of the semiconductor SPDC source and both single-photon detectors.
In the transceiver part, the probe photons are sent toward the target and the back-reflected probe photons are collected.
The two parts are connected via single-mode fiber, which enables a flexible deployment of the target detection system.
In the source and detection part, the photon pairs are generated through the SPDC process inside the semiconductor waveguide, which is externally pumped by a CW Ti-Sapphire laser at 783 nm.
The broadband, CW probe (with horizontal polarization) and reference(with vertical polarization) photons are separated upon a polarization beam-splitter.
The reference photons are coupled into a single-mode fiber and are detected on the reference detector.
The probe photons are coupled into another single-mode fiber and sent to the transceiver part.
In the transceiver part, the probe photons are emitted through an optical collimator towards the target object.
The probe photons back-scattered from the target object are collected by a telescope and are directed to the probe detector for detection.
To simulate strong environment noise, collimated broadband CW light shines towards the collection optics, leading to a large number of noise photons getting detected on the probe detector.
The noise light source is a light-emitting diode (BeST-SLQTCD\textsuperscript{\textregistered},LUMUX) with the output power attenuated by a tunable optical attenuator (HP\textsuperscript{\textregistered} 8156A). \\

\subsection{The target object and the collection optics}
The target object used in the experiment is aluminum foil, which has a high reflectivity but diffusive reflection at the frequency of the probe photons.
The target object is placed around 8 inches away from the first lens of the collection optics as shown in Fig. \ref{PIC}.
The presence and the absence of the target object are simulated by turning on and off of the coupling of the probe photon into the single mode fiber in the source and detection part, as shown in Fig. \ref{SETUP}.
This is carried out using a beam chopper to periodically block and unblock the probe photons at 2 Hz frequency with a 50:50 duty cycle.
The reason for not directly modulating the transmission of the probe photon in front of the target object is that doing so will affect the coupling of the noise photons into the collection optics as well, which could be due to the multiple reflections of the noise photons between the collection optics and the optical chopper.
Although physically different, turning on (off) the coupling of the probe photons within the source \& detector part is equivalent to the presence (absence) of the target object, both leading to the detection (non-detection) of the back-reflected probe photon at the probe detector.

The collection optics is a home-made telescope using three lenses.
The target object is placed around 8 inches away from the first lens of the telescope.
The collection efficiency, which determines the limit in our lab for target detection range, could be easily enhanced with an improved optical design of the telescope.
The total transmission of the probe photons \(\eta_p\) is estimated to be around 0.24\%, which includes various losses from the source to the detection, such as collection efficiency, detection efficiency, and the target reflectivity.
This value of \(\eta_p\) is obtained by conducting a trial experiment with long detection time and then the value of \(\eta_p\) could be calculated from experimentally measured photon detection statistics \(\mathbf{N}_r,\mathbf{N}_p,\mathbf{N}_c\) according to \eqref{PcEXPR}. %\eqref{Pp_expr} to \eqref{Pc_expr}.
The value of other experimental alparameters \(\nu,\eta_r,\nu_b\) could be calculated similarly.

\subsection{Single photon detectors}
The probe and reference detector are two superconducting nanowire single-photon detectors (Quantum Opus)\cite{QO}.
Both detectors are placed inside the same cryogenic system cooled down to 2.7K.
Both detectors have similar detection efficiency (\(\simeq 80\%\) at 1566 nm) and time uncertainty.
The time of each photon detection event on both detectors is recorded by a time-digital converter(ID800, IDQuantique).
The detector temporal uncertainty \(\Delta t\) (including the effect of electronic jittering of the voltage adapter between the detector and the time-digital converter and the time uncertainty of the time-digital converter) for both the reference and the probe detectors are 243 ps.\\
\begin{figure}[h]
\centering
\includegraphics[width=0.8\columnwidth]{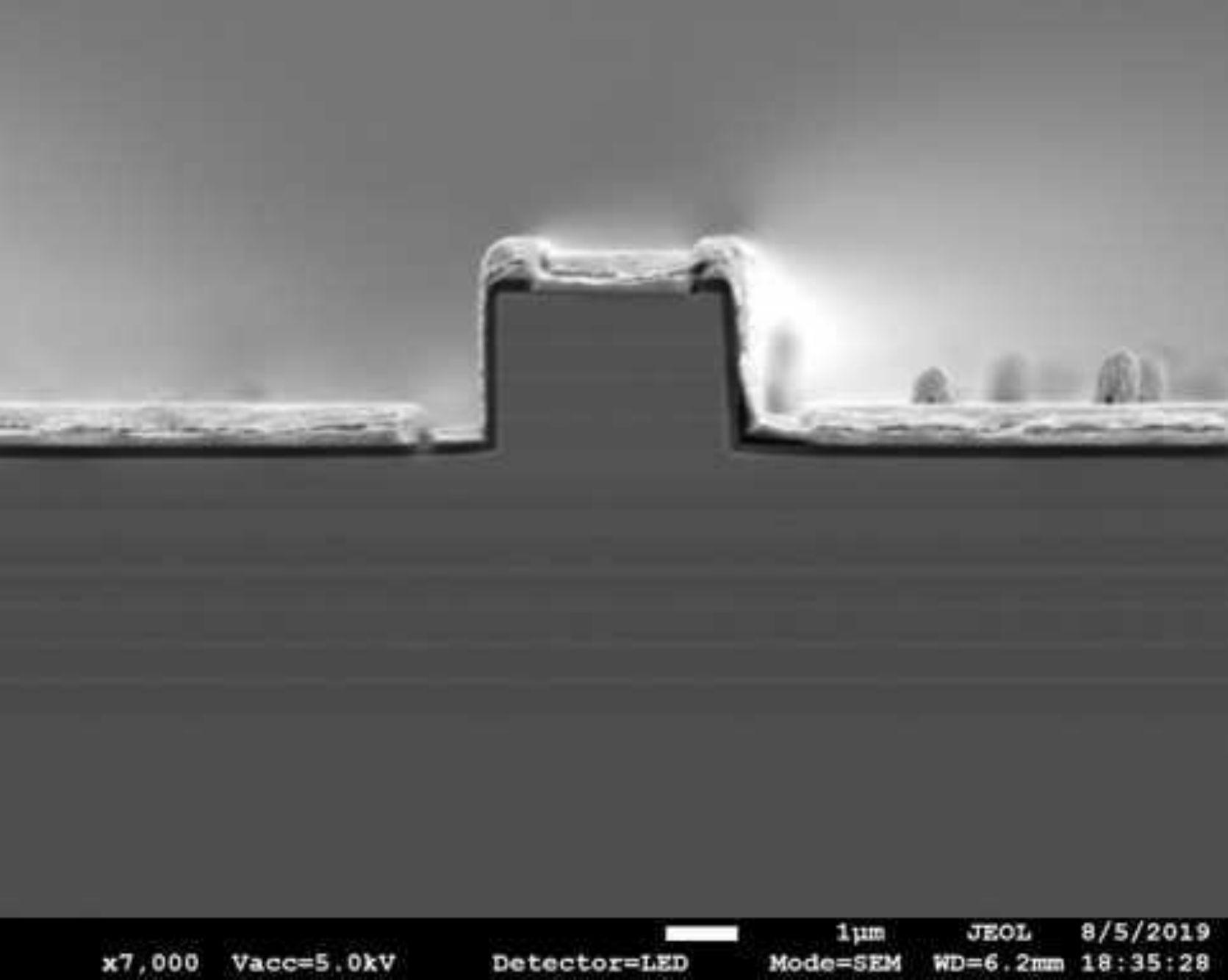}
\caption{the scanning electron microscope image of the semiconductor waveguide}\label{SEM}
\end{figure}

\subsection{Semiconductor waveguide source of non-classical photon pairs}
The semiconductor SPDC waveguide used is based on aluminum gallium arsenide (AlGaAs) material platform.
The dimension of the waveguide is \(\simeq 1\)mm long and \(\simeq 5\mu m\) wide.
The ridge and the substrate of the waveguide are based on the Bragg structure, consisting of multiple layers of AlGaAs alloy with different compositions (\(Al_xGa_{1-x}As,0\le x \le1\)) as could be seen in Fig. \ref{SEM} \cite{Horn:2012}.
This semi-periodical structure of the waveguide provides simultaneous confinement of the 783nm pump light and the generated photon pairs around 1566 nm.
In addition, the waveguide structure is also designed to satisfy the momentum matching condition\cite{boyd2003nonlinear} of the SPDC process.
The conversion efficiency is estimated to be 2.1\(\times 10^{-8}\) probe-reference photon pairs per 783nm pump photon.
In the experiment setup, the nonlinear waveguide is mounted on a two-dimensional translation stage.
The input 783nm pump light is coupled into the rear facet of the waveguide through an objective lens.
The probe-reference photon pairs are coupled out from the front facet waveguide through another objective lens.
\\

The advantages of the semiconductor photon-pair source for target detection as compared to conventional SPDC source based on bulk nonlinear crystals are many-fold.
\begin{itemize}
\item The significantly smaller form factor of the semiconductor waveguide enables large scale integration of photon pairs sources, which is crucial for realistic target detection applications where a large flux of non-classical photon pairs is needed.\\

\item For each of the waveguides, the generation of the probe-reference photon pairs is also efficient because of the strong nonlinearity of the semiconductor material.\\

\item The AlGaAs platform also allows the possibility of the active waveguide with electrically pumping (i.e. generating the correlated photon pairs by applying a voltage across the waveguide), which is also favorable in terms of large scale integration.\\

\item The spectrum of the SPDC photon pairs could be engineered through different designs of the waveguide structure\cite{Abolghasem:2009}. For example, the central frequency of the probe could be shifted to suit the need in different target detection scenarios without shifting the central frequency of the reference photons.\\
\end{itemize}
\begin{figure*}[!h]
\centering
\includegraphics[width=0.8\columnwidth]{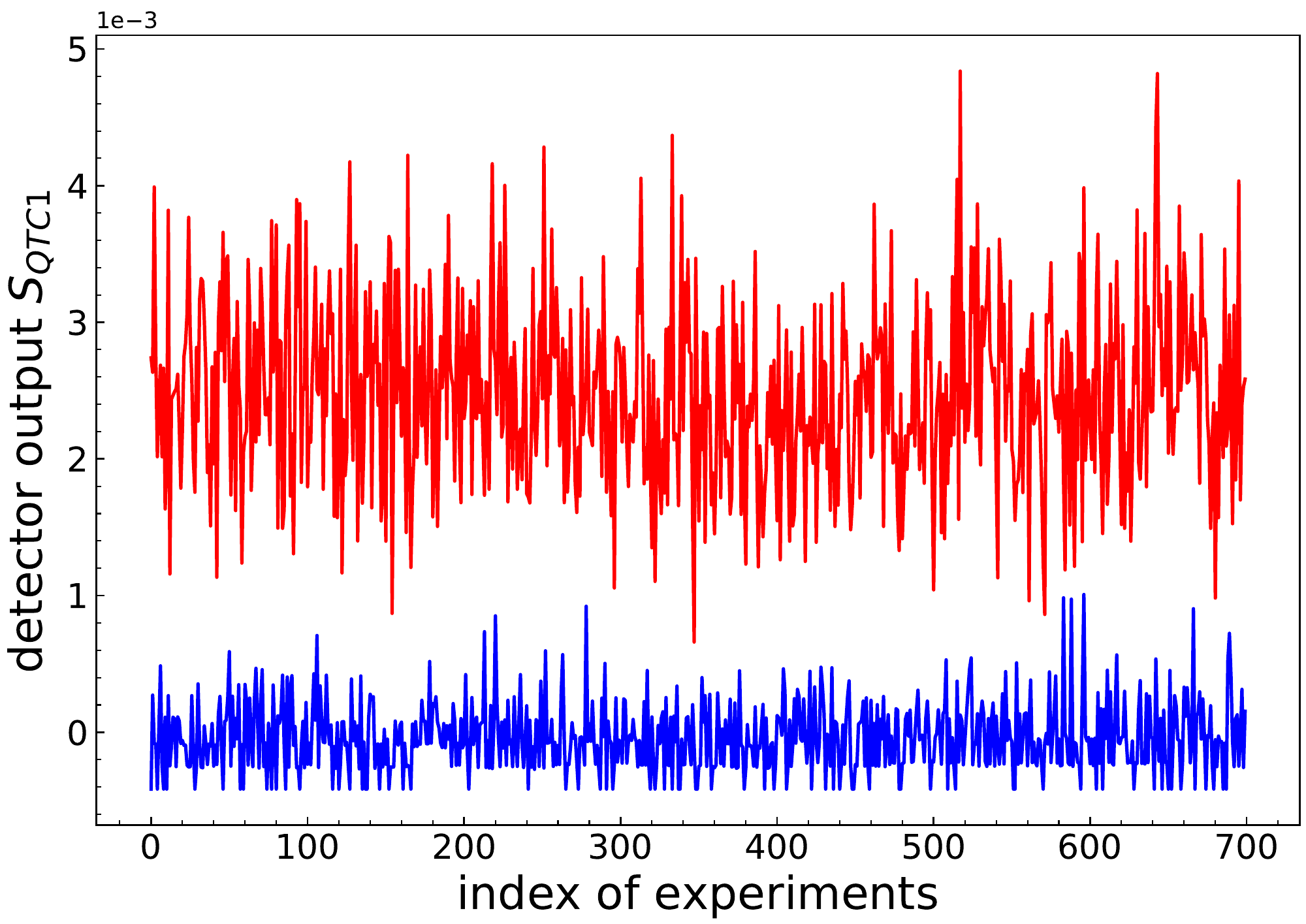}\hspace{1cm}
\includegraphics[width=0.8\columnwidth]{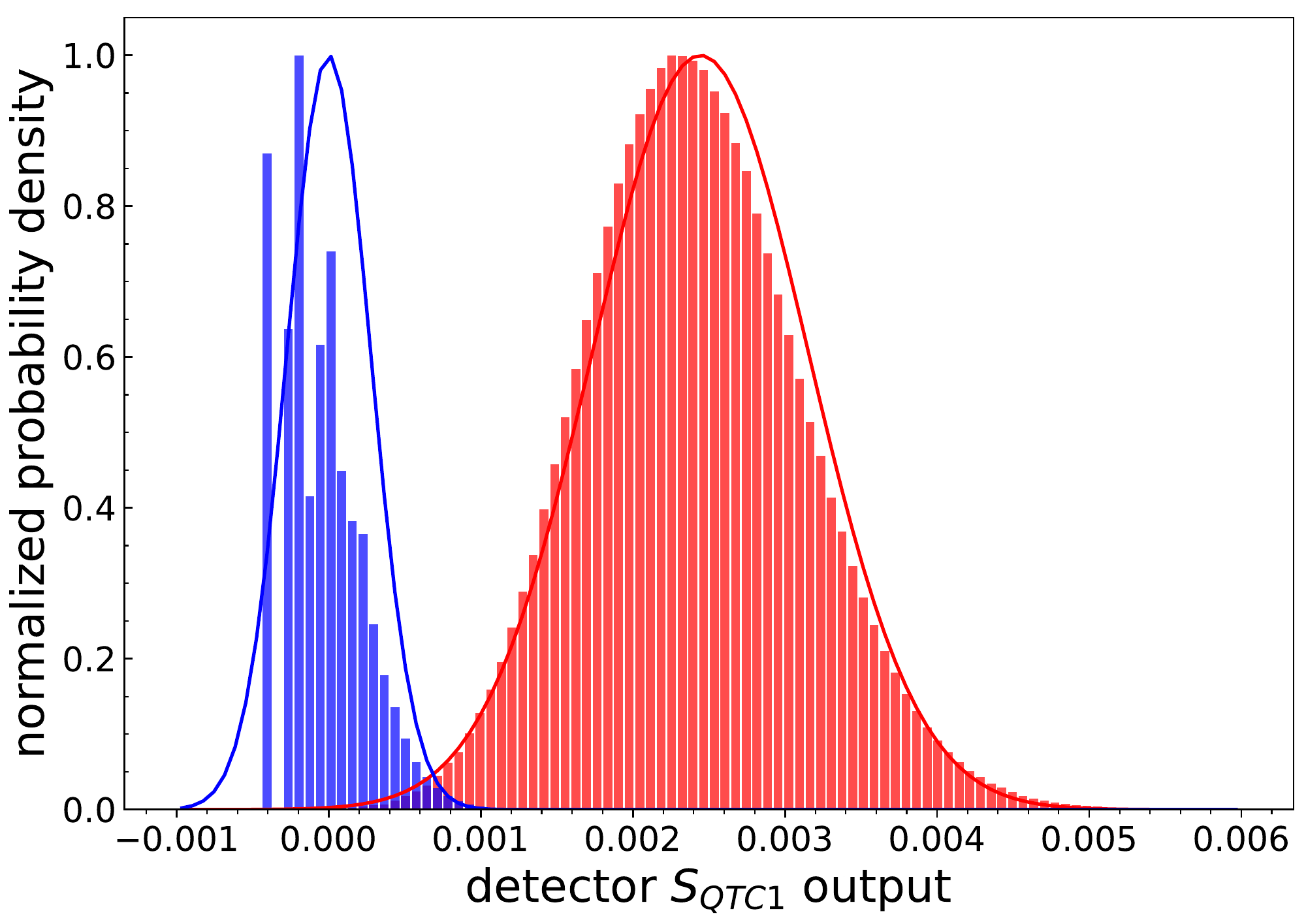}\\
\includegraphics[width=0.8\columnwidth]{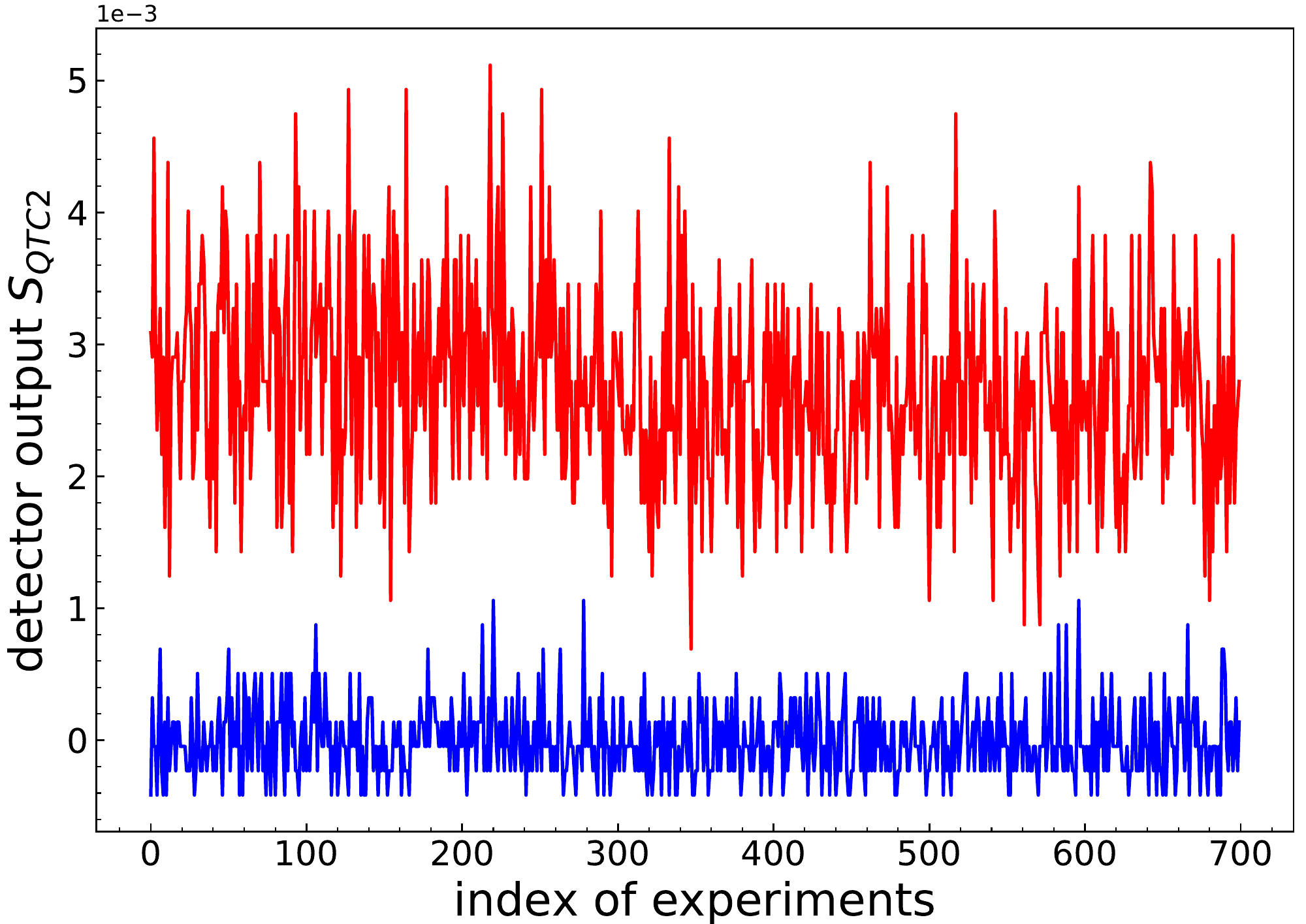}\hspace{1cm}
\includegraphics[width=0.8\columnwidth]{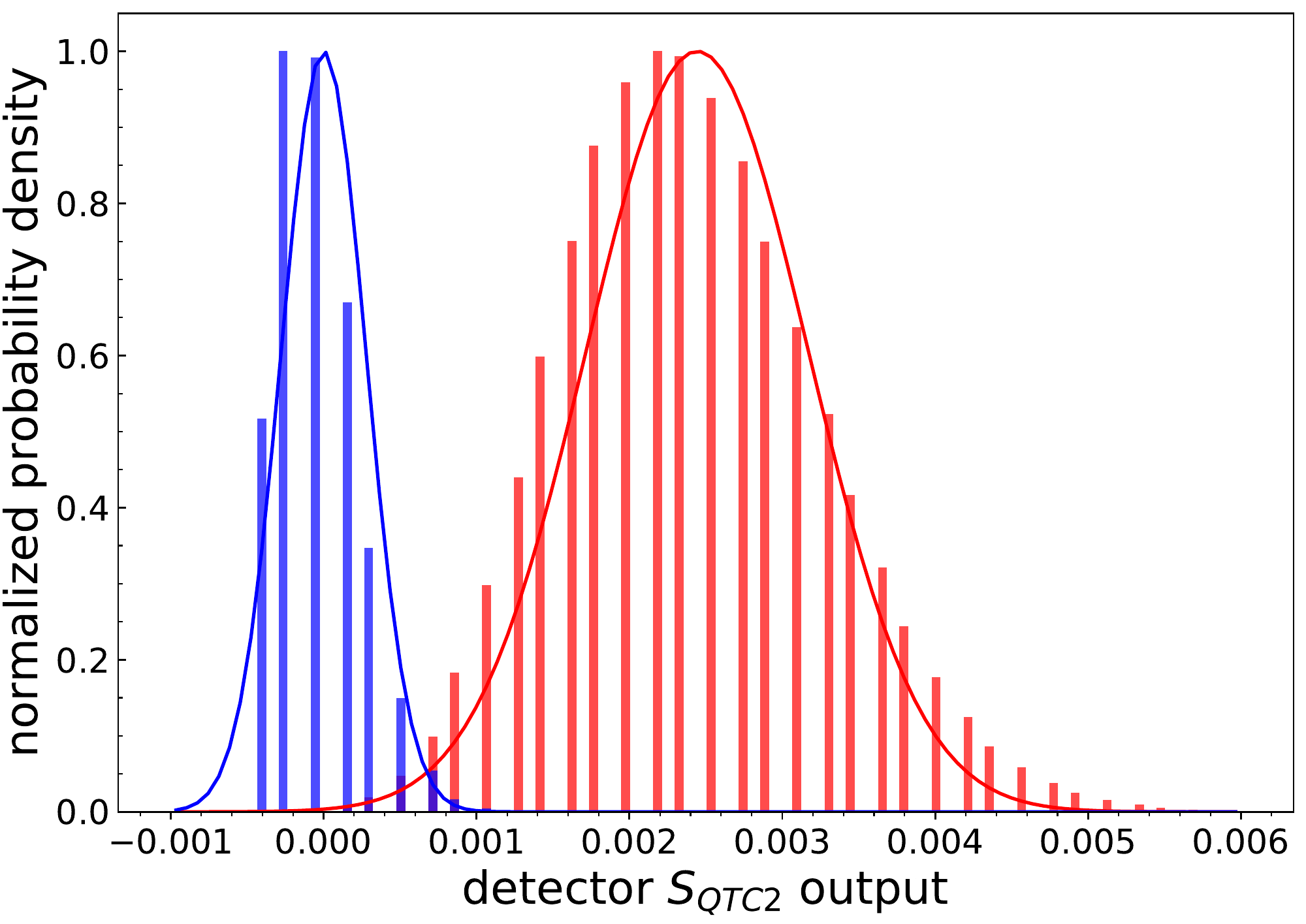}\\
\includegraphics[width=0.8\columnwidth]{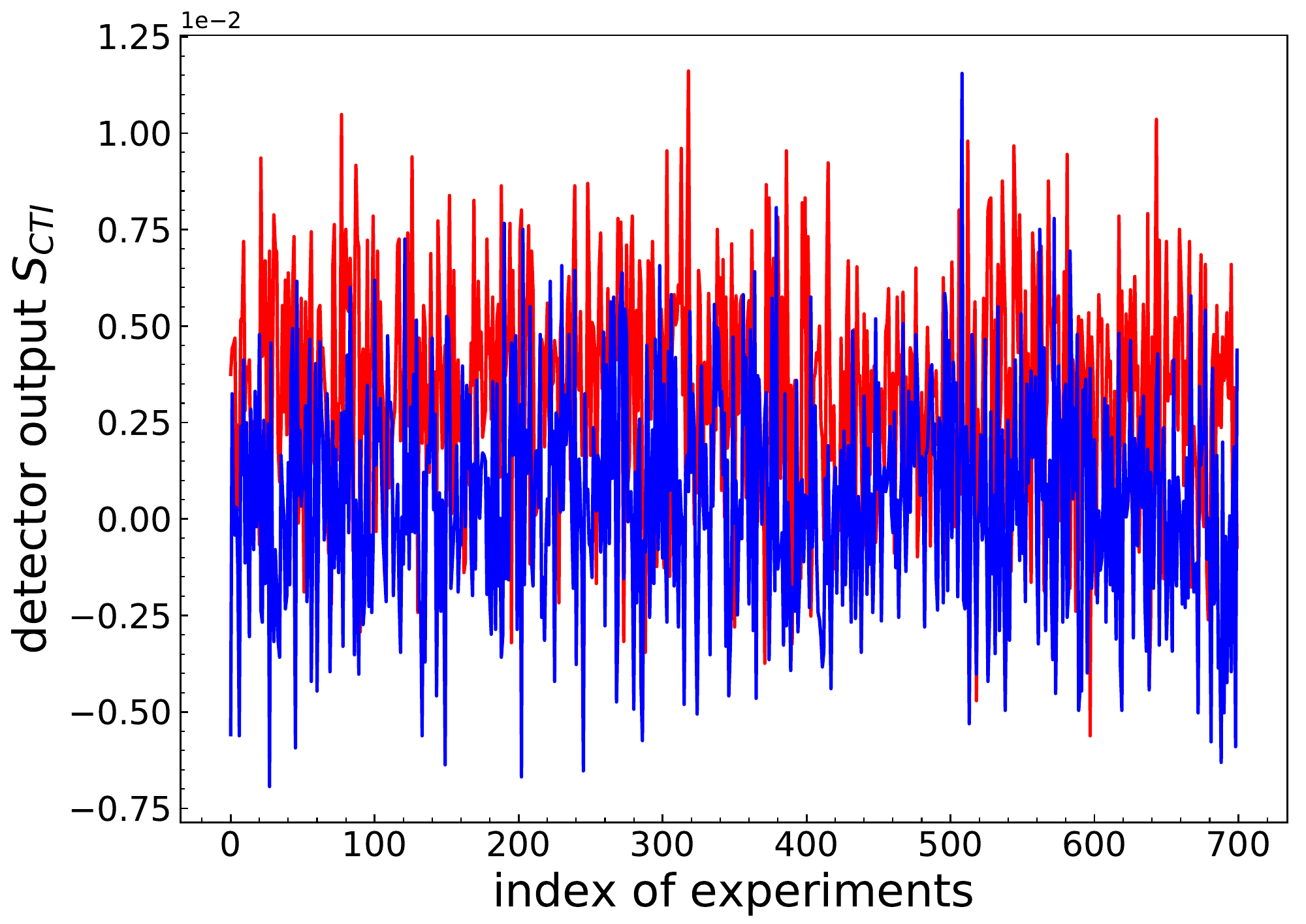}\hspace{1cm}
\includegraphics[width=0.8\columnwidth]{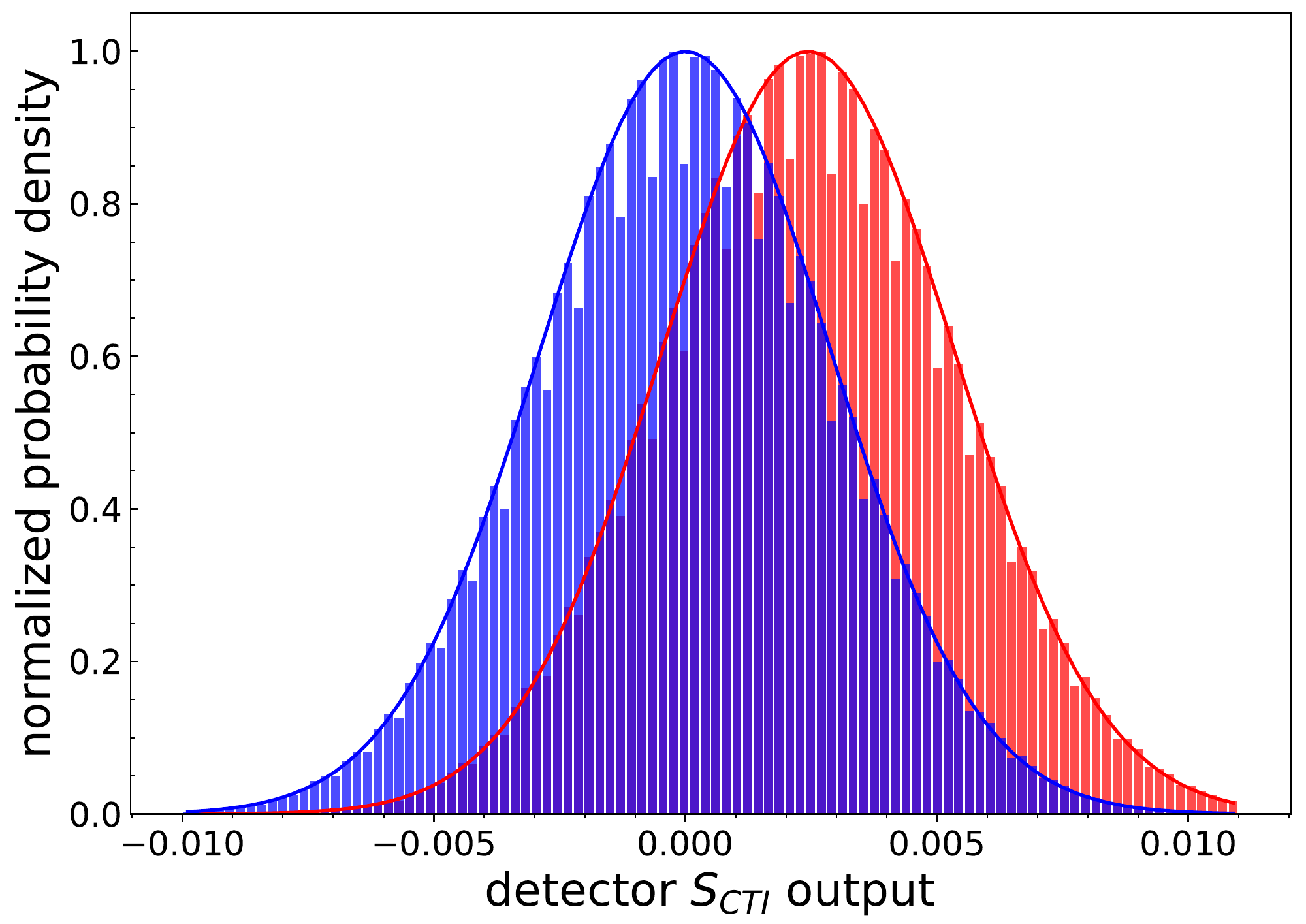}\\

\caption{Left:the time series of different detector functions \(\mathbf{S}_\text{QTC1}\), \(\mathbf{S}_\text{QTC2}\) and \(\mathbf{S}_\text{CTI}\) over 1400 independent target experiment.
X-axis: the index of independent experiment that last time \(\tau=0.01\)s.
Blue (red) bar: probability density when the target is absent (present).
Blue (red) solid line, the theoretically calculated probability distribution using the Gaussian signal approximation \eqref{GAPPROX}.
The histograms are calculated from a larger portion of data (\(2.19\times10^{6}\) independent experiments).
}\label{TS_HIST}
\end{figure*}
\begin{figure*}[!h]
\centering
\includegraphics[width=0.8\columnwidth]{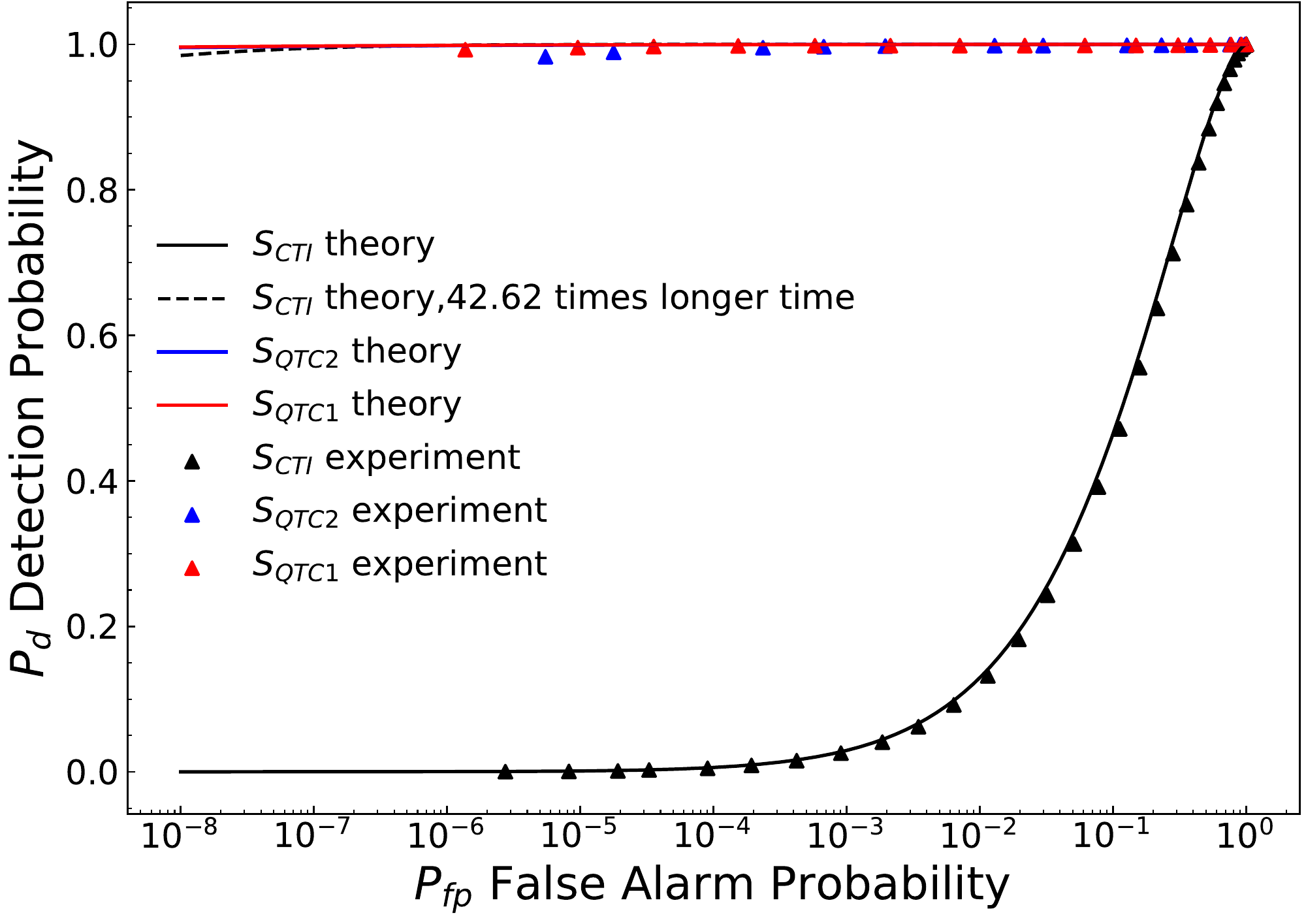}\hspace{1cm}
\includegraphics[width=0.8\columnwidth]{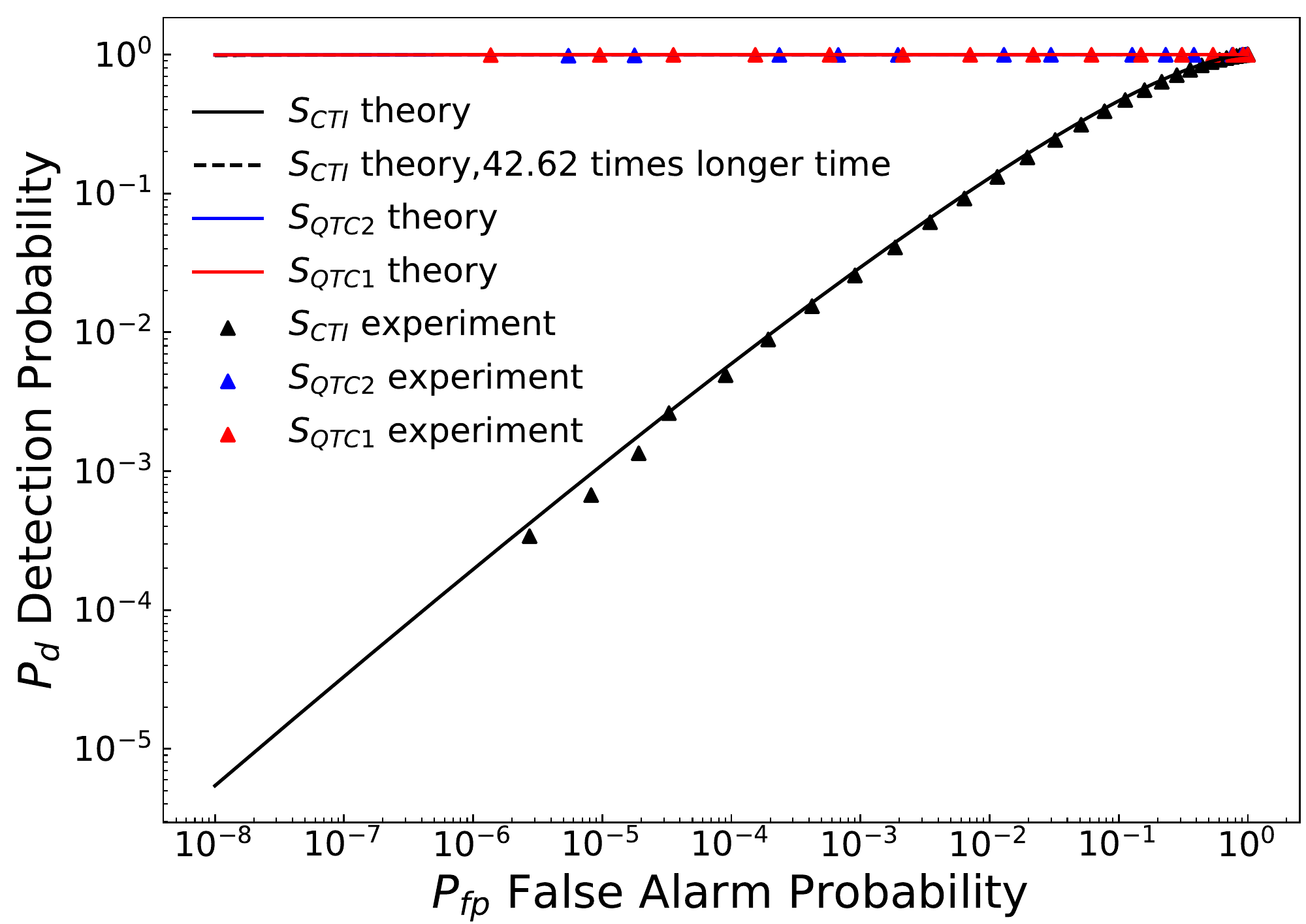}\\
\includegraphics[width=0.8\columnwidth]{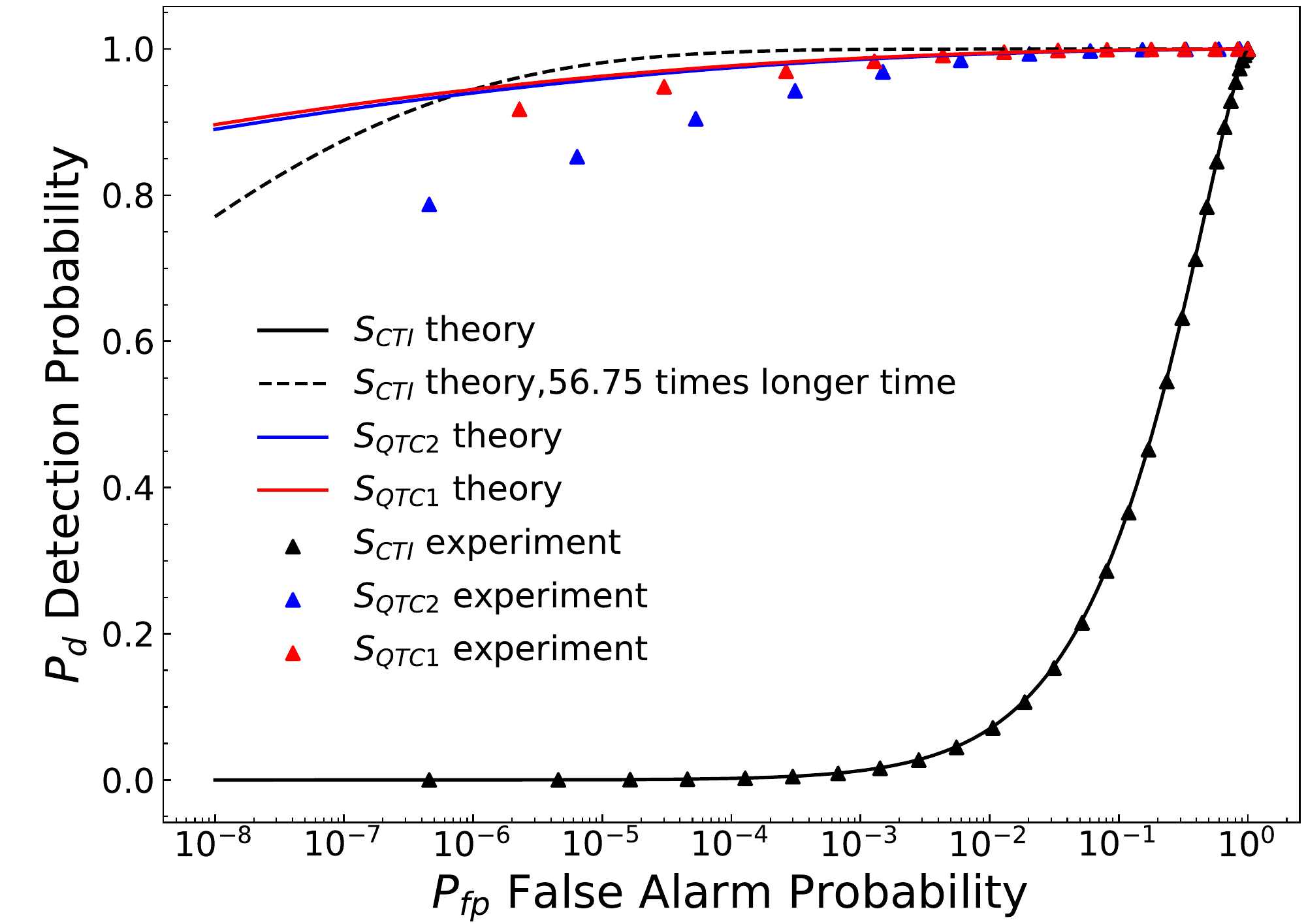}\hspace{1cm}
\includegraphics[width=0.8\columnwidth]{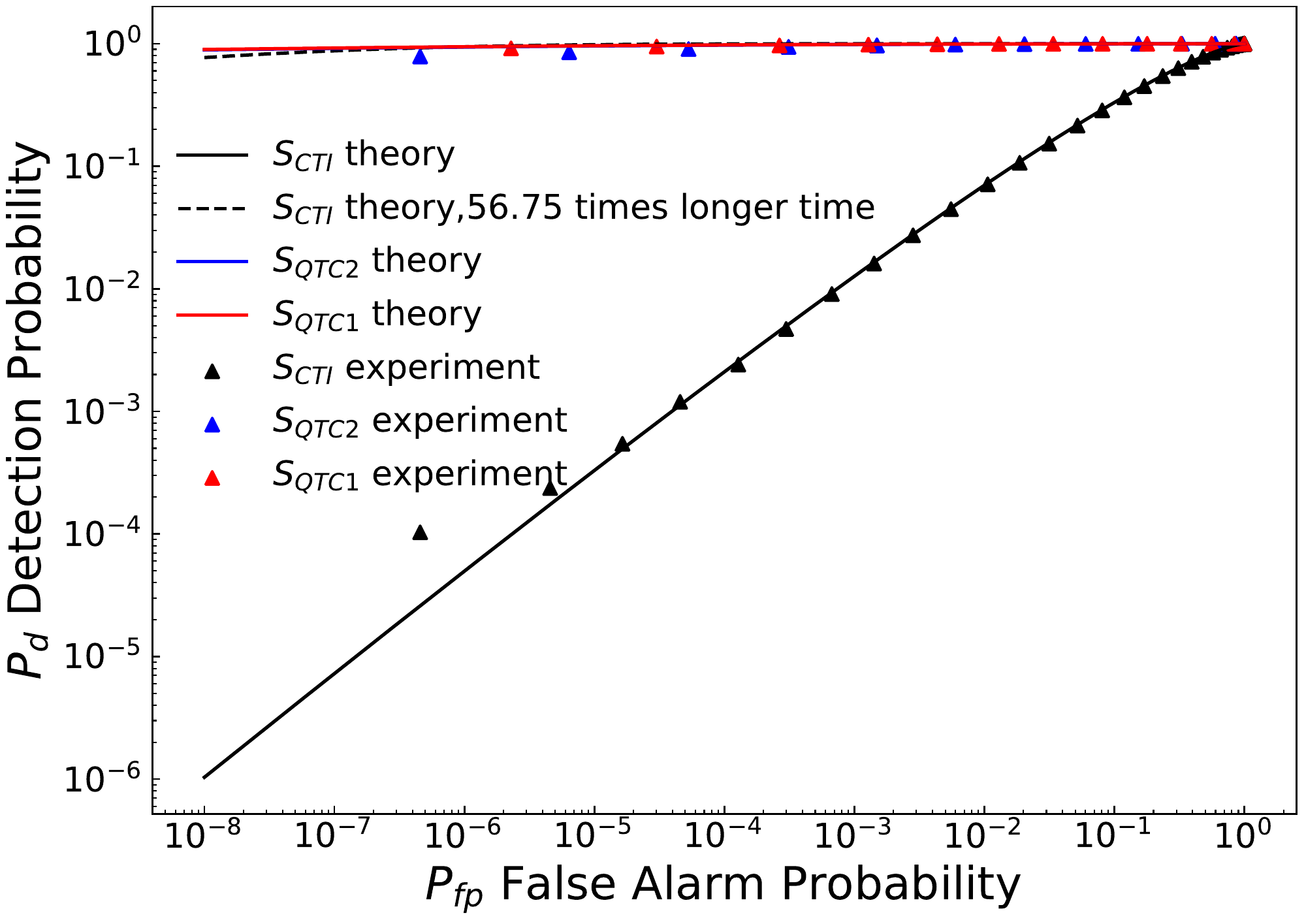}\\
\includegraphics[width=0.8\columnwidth]{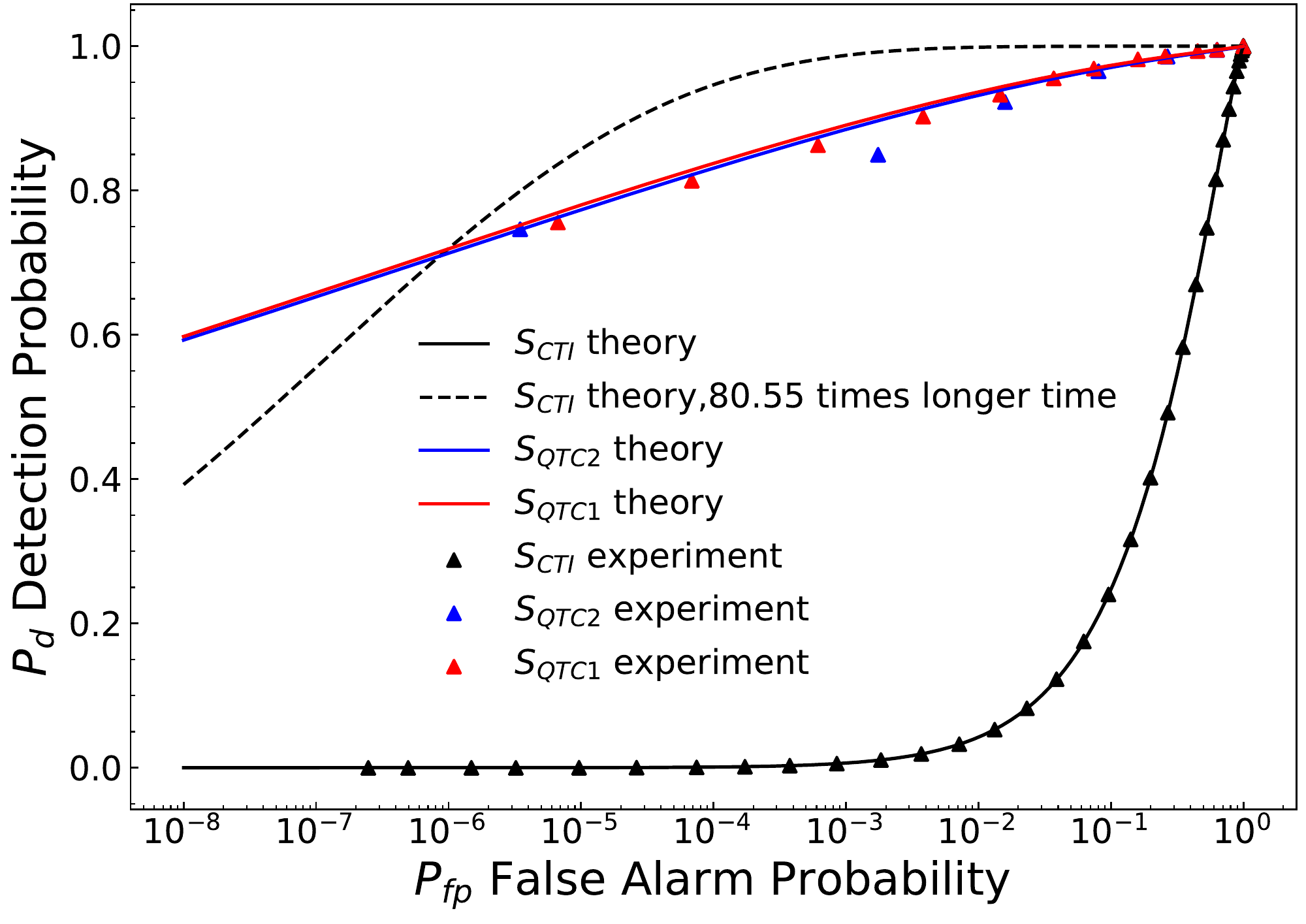}\hspace{1cm}
\includegraphics[width=0.8\columnwidth]{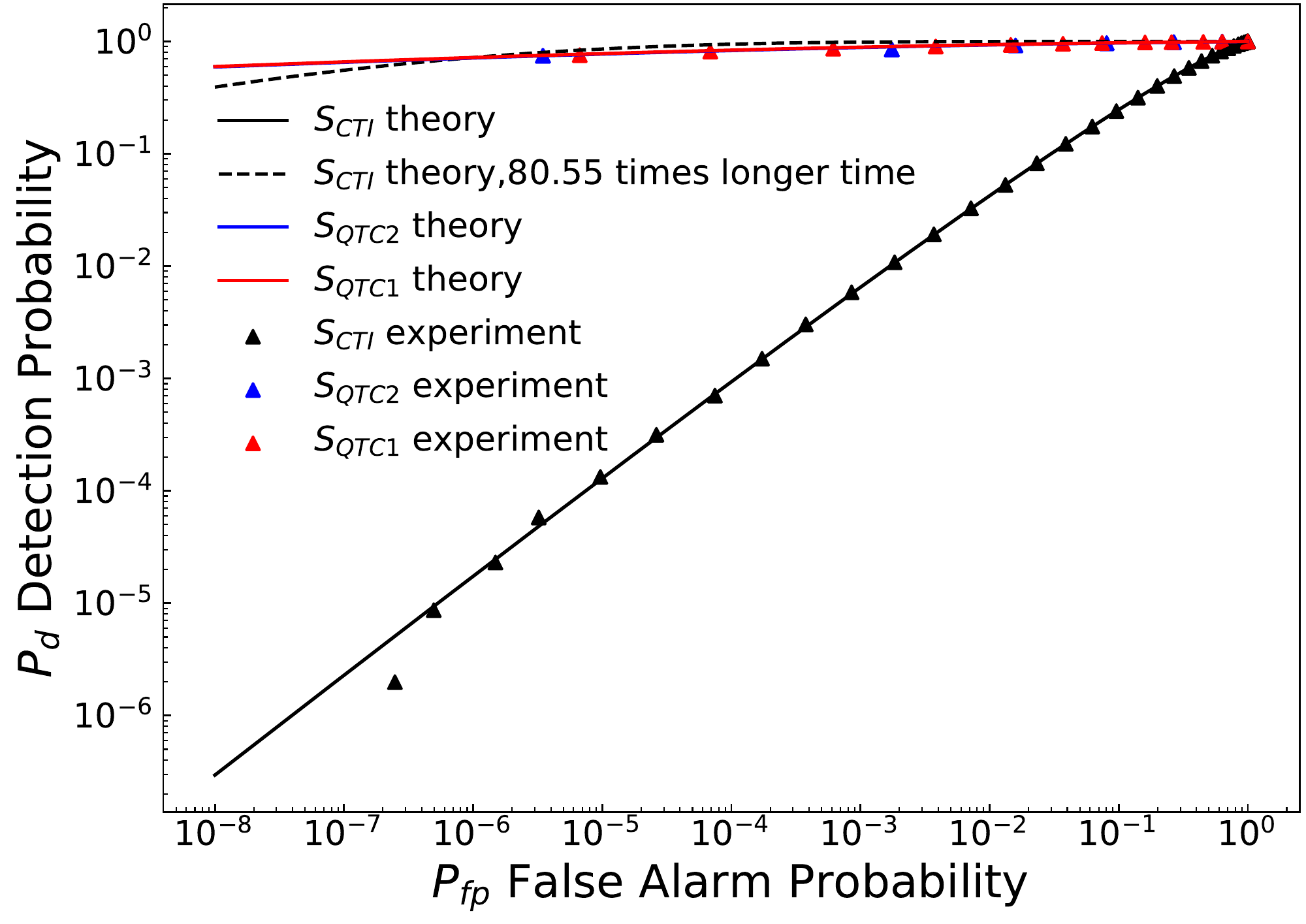}
\caption{Left:ROC curves for the detector function \(\mathbf{S}_\text{QTC1}\), \(\mathbf{S}_\text{QTC2}\) and \(\mathbf{S}_\text{CTI}\) for \(\tau=0.02\)s, \(\tau=0.01\)s and \(\tau_=0.005\)s (from top to bottom).
Red curve: the theoretical ROC curve for \(\mathbf{S}_\text{QTC1}\) calculated based on the Gaussian signal approximation.
Red triangles:the experimental ROC curve for \(\mathbf{S}_\text{QTC1}\).
The blue curve, blue triangles, and black curve, black triangles are similarly plotted for detector functions \(\mathbf{S}_\text{QTC2}\) and \(\mathbf{S}_\text{CTI}\), respectively.
Dashed black curve: theoretical ROC curve for the detector function \(\mathbf{S}_\text{CTI}\) with longer integration time \(\tau'=k_\tau\tau\) such that the detection rate \(P_d\) same as that of \(\mathbf{S}_\text{QTC1}\) is achieved at false alarm rate \(P_{fa} =10^{-6}\). The ratios are \(k_\tau=42.62,56.75,80.55\) for \(\tau=0.005s,0.01s,0.02s\). For the same integration time \(\tau\), the CTI detection protocol can also achieve the same performance with \(6.59,7.59,9.01\) times more source power or total probe photon transmission.
Right: the same ROC curve plotted with log scaled \(P_d\) axis.
}\label{ROC_FIG}
\end{figure*}

\section{Experimental results}
We conduct repeated (\(\simeq 2\times10^6\)) target detection experiments for the QTC detection protocol to characterize its performance.
Each of the experiments lasts for a short period \(\tau=0.01s\).
The target object is absent (probe photons blocked) for half of the total number of the experiments.
The same experimentally recorded photon counting data is used for the CTI detection protocol too, but the photon detection events on the reference detector are neglected.
By doing so the drift of the experiment condition between the CTI detection protocol experiment and the QTC detection protocol experiment is eliminated.
A table of experimental parameters could be found in the table(\ref{PARAMS})(note that the ROC curves are also plotted for different values of measurement time \(\tau\) apart from the listed value). A qualitative sense of the experiment is provided by the ratio of the number of probe photons and the number of noise photons, i.e., $10\log_{10}(\nu \eta_p/\nu_b)\approx-20$ dB.
The performance of the three detector functions \(\mathbf{S}_\text{QTC2}\), \(\mathbf{S}_\text{QTC1}\) and \(\mathbf{S}_\text{CTI}\) are quantified and compared through the ROC analysis of the experimental photon counting data.

\subsection{Time series, histogram and detector performances}
The experimental photon counting data is recorded by the ID800 time-digital converter in a `detection time - channel index' format, where the `channel index' stands for either the probe or the reference detector.
A coincidence detection event is registered if two consecutive photon detection events are on different detectors and have time difference shorter than half the coincidence window \(\frac{T_c}{2}\) (after compensating the difference of the optical path lengths of the probe and reference photon).
The number of single-channel detection events \(\mathbf{N}_p\) and \(\mathbf{N}_r\) are obtained by subtracting the total number of coincidence detection events \(\mathbf{N}_c\) from the total number of photon detection events on the probe and the reference detector.
From these statistics different detector functions \(\mathbf{S}_\text{QTC1}\), \(\mathbf{S}_\text{QTC2}\) and \(\mathbf{S}_\text{CTI}\) could be evaluated.
Fig. \ref{TS_HIST} shows the time series of the three detector functions from 1400 repetitions of independent target detection experiments (700 experiments with the target object present and 700 experiments with the target object absent).
As could be seen, both detector functions for the QTC detection protocol (\(\mathbf{S}_\text{QTC1}\) and \(\mathbf{S}_\text{QTC2}\)) have much lower fluctuation compared to the detector function of the CTI detection protocol.
The histogram of the values of the different detector functions is also shown in Fig. \ref{TS_HIST}.
The experimentally measured histograms agree well with the theoretical theory curve that is obtained from the Gaussian signal approximation, which validates such approximation.
For the detector function \(\mathbf{S}_\text{QTC2}\) there exists 'gaps' within the histogram.
This is because the number of coincidence detection events is relatively low and quantized in each experiment.

\subsection{ROC curve}
The ROC curves of the different detector functions are plotted in Fig. \ref{ROC_FIG} with different value of experiment time \(\tau=0.005,0.01,0.02\)s.
For the CTI detection protocol, the experimental ROC curve perfectly matches the theoretical ROC curve.
The theoretical ROC curve of detector function \(\mathbf{S}_\text{QTC2}\) is only slightly lower than that of the \(\mathbf{S}_\text{QTC1}\) detector, suggesting that coincidence detection is close to optimal for the QTC detection protocol.
The experimental ROC curve for \(\mathbf{S}_\text{QTC1}\) and \(\mathbf{S}_\text{QTC2}\) detector functions are considerably lower than the theoretical prediction.
We attribute this to the drift of the SPDC pump power and the probe transmission \(\eta_p\) over time.
This is confirmed as the drift of the SPDC pump power and the probe transmission is experimentally observed. Moreover, when the ROC curves are plotted for a smaller sample of independent experiments that last a shorter duration, the theoretical and experimental curve will have a better matching.
The fact that the experimental ROC curve for the detection function \(\mathbf{S}_\text{QTC1}\) is considerably higher than that of \(\mathbf{S}_\text{QTC2}\) may suggest the better robustness of the \(\mathbf{S}_\text{QTC1}\) against the drift of the experiment condition.\\

Both detector functions \(\mathbf{S}_\text{QTC1}\) and \(\mathbf{S}_\text{QTC2}\) have significant performance advantages over the CTI detection protocol detector function.
For measurement time \(\tau=0.01\)s, when the false alarm rate is set around \(P_{fa}=10^{-6}\), the experimental detection rate of \(\mathbf{S}_\text{QTC1}\) and \(\mathbf{S}_\text{QTC2}\) are around \(80\%\).
To achieve the same theoretical detection rate \(P_d\) of the detector function \(\mathbf{S}_\text{QTC1}\), the CTI detection protocol must last for around \(\simeq57\) times longer, as shown in Fig. \ref{ROC_FIG}.
Such a time reduction factor implies the much higher target detection speed of the QTC detection protocol.
Alternatively, for the same measurement time \(\tau=0.01\), the CTI achieve can achieve the same detection rate \(P_d\) with \(\simeq\)7.6 times more probe power or total probe channel transmission.
Fig. \ref{ROC_FIG} also shows the different ROC curves with different measurement time \(\tau\). It could be seen that the QTC detection protocol could have even higher performance advantages compared to the CTI detection protocol when the detection time \(\tau\) is short and when the false alarm rate \(P_{fa}\) is low.

\begin{figure}[ht]
\centering
\includegraphics[width=0.8\columnwidth]{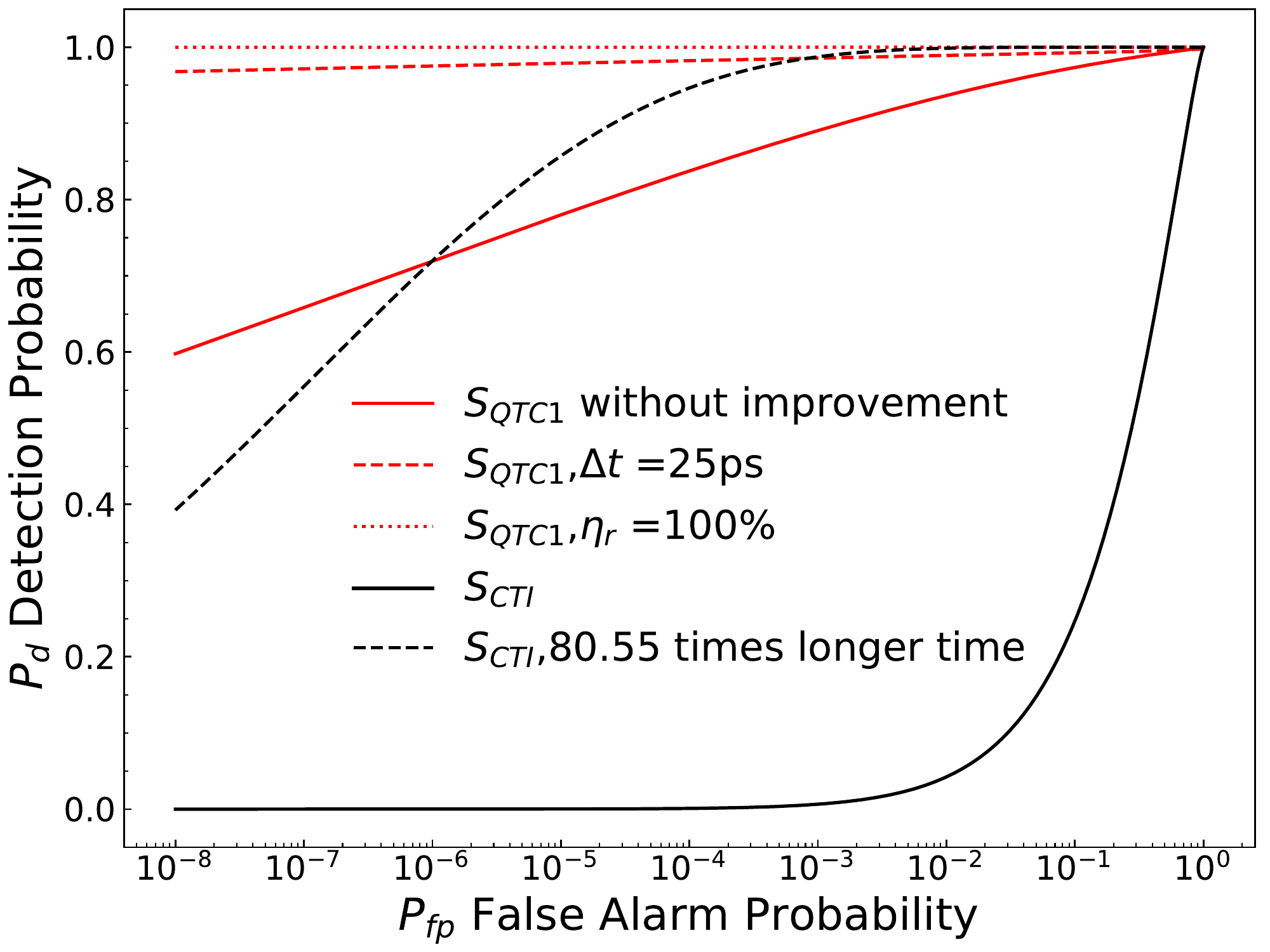}
\caption{Theoretical ROC curves with measurement time \(\tau=0.005\)s and improved effective time uncertainty \(\Delta t_{eff}\) or reference photon transmission \(\eta_r\) for the QTC detection protocol. Black: the CTI detection protocol for comparison; Red: the QTC detection protocol (\(\mathbf{S}_\text{QTC1}\)) without any improvement; Black dashed: the CTI detection protocol with \(\simeq80.55\) times longer measurement time; Red dashed: the QTC detection protocol with improved effective time uncertainty \(\Delta t_{eff}=25 ps\); Red dotted: the QTC detection protocol with perfect reference photon transmission \(\eta_r=100\%\)} \label{ROC_FIG_OPT}
\end{figure}

\section{Discussion}

\subsection{Factors impacting the target detection performance}
The factors influencing the target detection performance could be classified into two sets:
\begin{itemize}
\item  Factors that only affect the performance of the QTC detection protocol, namely, reference photon transmission efficiency ($\eta_r$) and effective time uncertainty ($\Delta t_{eff}$);
\item Factors that affect both the QTC detection and the CTI detection protocols, particularly,  the source pair rate  ($\nu$), background noise photon detection rate ($\nu_b$), and probe photon transmission efficiency ($\eta_p$).
\end{itemize}
 The most important limiting factor that only affects the QTC detection protocol is the transmission efficiency,  \(\eta_r\simeq20\%\) in our experiment, of the reference photon.
In the limit of zero reference photon transmission \(\eta_r = 0\), the QTC detection protocol will be identical to the CTI detection protocol and hence provide no advantage.
In the experiment, \(\eta_r\) is mainly limited by the finite efficiency of the coupling of the reference photon into the single-mode fiber that is connected to the reference detector.
This could be due to the low modal overlap between the waveguide mode and the fiber mode for the reference and probe photons.
Such inefficiency could be alleviated with an improved design of the coupling optics, or the semiconductor waveguide to achieve better modal overlap.
In the limit of perfect reference photon transmission (\(\eta_r = 100\%\)) the ROC curve of the QTC detection protocol ( with the detector function \(S_\text{QTC1}\)) is plotted in Fig. \ref{ROC_FIG_OPT}.

Another important limiting factor for the QTC detection protocol only is the effective time uncertainty \(\Delta t_{eff}\), which determines the temporal coincidence window \(T_c=2\Delta t_{eff}\).
From \eqref{FI_EXPR_NC} and \eqref{PcEXPR} it could be seen that for the detector function \(S_\text{QTC2}\)(whose performance is close to the detector function \(S_\text{QTC1}\)), the effect of background noise could be reduced by a factor of \(x\) when the effective time uncertainty \(\Delta t\) is reduced by a factor of \(x\).
The effective time uncertainty \(\Delta t_{eff}\) is limited by the intrinsic correlation time \(\Delta t_0\) and the the detector time uncertainty \(\Delta t\).
In our current implementation of the QTC detection protocol the effective temporal uncertainty \(\Delta t_{eff}\) is mainly limited by the detector temporal uncertainty \(\Delta t_{eff} \simeq \Delta t  =243\) ps.
This value could be improved with single-photon detectors that have a faster response or with improved photon detection techniques \cite{liu2019enhancing}.
Fig. \ref{ROC_FIG_OPT} also shows the ROC curve of the QTC detection protocol for the detection function \(S_\text{QTC1}\) when the effective time uncertainty is reduced to \(\Delta t_{eff}\simeq 25\) ps.\\

Factors affecting both the QTC detection protocol and the CTI detection protocol performance include the source pair rate \(\nu\), background noise photon detection rate \(\nu_b\) and the probe photon transmission efficiency \(\eta_p\). The performance of both the QTC detection protocol and the CTI detection protocol increase when the source pair rate \(\nu\) is high and the background noise power \(\nu_b\) is low. This corresponds to the intuitive idea that better target detection performance could be achieved with higher source power and lower background noise. However, it is worth noting that \(\nu\) and \(\nu_b\) does not affect the QTC and the CTI detection protocols equally. The performance advantage of the QTC detection protocol over the CTI detection protocol is larger in high noise \(\nu_b\) and low source power \(\nu\) regime\cite{liu2019enhancing}, suggesting that the QTC detection protocol is more favorable for target detection with low power and high noise background.\\

The total probe transmission efficiency, \(\eta_p\), impacts the performance of both the QTC detection protocol and the CTI detection protocol. The total transmission \(\eta_p\) is affected by many factors including the target distance, the reflectivity of the target, propagation medium, and the probe photon collection efficiency.
This corresponds to the intuitive idea that target within a short distance and high reflectivity is easy to detect.
The experiment result and theoretical analysis show the significant performance advantages of the QTC detection protocol over the CTI detection protocol with the same \(\eta_p\). In particular, the variance of the estimation can be considerably lower for the QTC detection protocol.
This suggests that for the same level of target detection performance, the QTC detection protocol is capable of detecting a target object with lower reflectivity and longer distance than the CTI detection protocol.

\subsection{QTC and CTI detection protocols with pulsed or narrowband probe}

It is important to note that in the theoretical model and the experiment of both the QTC and CTI detection protocol, the background noise is assumed to be overlapping with the probe photon in both temporal and spectral domains, i.e., broadband and CW.
Therefore is not possible to reduce the in-band noise power through filtering.
However, if a short pulse or narrowband probe light is utilized, then it is possible for the CTI detection protocol to reduce the in-band noise power through spectral filtering or temporal gating.
In such cases, the performance of the CTI detection protocol could be increased and is also ultimately limited by the detector time or frequency uncertainty.
Nevertheless, the utilization of short-pulse or narrow-band probe light will not only increase the system complexity by using a pulsed or narrowband laser but also reduced the stealth of the target detection:
such noise reduction comes at the price of concentrated optical power in either the temporal or spectral domain, and hence increasing the visibility of the target detection channel.
If the adversary party is able to distinguish the probe photons from the background noise by their different spectral-temporal properties, then they could selectively jam the target detection channel with noise photons that are indistinguishable from the probe photons.
Although there exist classical scrambling techniques such as frequency scrambling to increase the indistinguishability between the probe and noise photons, such indistinguishability is not guaranteed by fundamental physical principles.
On the other hand, the QTC detection protocol provides unconditional indistinguishability between the probe and noise photons.
This is because each of the probe photons is generated at a fundamentally random time and frequency, albeit its strong temporal correlation to the reference photon.
When the environment noise power is very high compared to the probe power, the probe photon will be almost invisible to the adversary party, but the QTC detection protocol could still achieve meaningful target detection performance, as shown in the results and analysis above.\\

Another important advantage of the QTC detection protocol is jamming resilience. Note that in the case of pulsed source, if the background photons and probe photons arrive at the same time, it would be totally and perfectly jammed. However, intuitively, the QTC detection protocol based on CW pumping is more resilient to jamming noise as the probe photons arrive at random times.  This intuition is confirmed by the theoretical analysis of the Fisher information.

\subsection{Covert ranging}
Covert ranging is an application of the QTC detection protocol that could take advantage of its stealth property.
Classical ranging protocol typically utilizes pulsed electromagnetic radiation to be probe the target, and the target distance could be calculated from the time-of-flight of the probe photon.
However, time-variant probe radiation is distinguishable from the CW background noise and is therefore visible to the unauthorized receivers.
Ranging with classical time-invariant radiation is not possible since the time-invariant back-reflected signal does not contain much information about the target distance.
On the other hand, ranging with time-invariant probe light is possible for the QTC detection protocol.
Since each probe photon is temporally correlated with a reference photon, the travel distance of the probe photon could be calculated from the detection time difference of the probe and reference photon.
Meanwhile, the probe photon in the QTC detection protocol is generated at a fundamentally random time and frequency and is therefore indistinguishable from the CW broadband environment noise.
A preliminary ranging experiment with the QTC detection protocol was reported in \cite{liu2019enhancing}, achieving distance resolution of \(\simeq\)5 cm, which is limited by the detector time uncertainty.

\subsection{Other enhanced target detection protocols}
The major difference between the QTC detection protocol, discussed here, and the previously reported quantum two-mode squeezing (QTMS) detection protocol \cite{chang2019quantum} is the different type of correlation of the photon pairs that is utilized.
In the QTMS protocol, which is operating in radio frequency, the phase-sensitive complex amplitude correlation between the probe and reference light is utilized.
Correspondingly, the detection in the QTMS protocol consists of phase-sensitive heterodyne detection of both the reference and the probe photons.
In the optical frequency, the generation of complex amplitude correlated state (squeezed state) is also possible through spontaneous parametric down-conversion process \cite{vahlbruch2016detection}.
However, the homodyne measurement that is required to measure such complex amplitude correlation is much more challenging.
In optical frequencies, homodyne measurement requires precise temporal matching of the waveform of the local oscillator and the probe/reference photon, which means the spatial overlap between the local oscillator and the probe light have to be stabled to sub-wavelength level.
Such stringent requirements make phase-sensitive homodyne measurement impractical for optical target detection and ranging since the exact position of the target is usually unknown or unstable.\\

Compared to the phase-sensitive heterodyne measurement that is used in QTMS radar, time-resolved photon detection in the QTC detection protocol is phase-insensitive and easy to implement in the optical domain. This is because single-photon detection technologies are relatively mature in optics, while quadrature (I and Q) measurements are very challenging.
It is worth emphasizing that this does not mean the QTC detection protocol is restricted to the optical frequencies.  In fact, the relatively higher noise background in the microwave regime could potentially make the advantages of the QTC detection protocol over the classical intensity detection based protocol even more evident. However, it must be noted that in the microwave regime,  one can do better using coherent signal processing (e.g., match filtering), and that current radars do not use match filtering in the single-photon level. So a more detailed analysis needs to be carried out to investigate utility of QTC in the microwave regime.

In the optical domain, there are other correlation-enhanced target detection protocols utilizing the photon-number correlation in discrete timebins\cite{lopaeva:2013, BBDE2018}. These protocols are not suitable for covert operations since they utilized pulsed photon pairs as the sources that are distinguishable from the CW noise background. Moreover, the bulk crystal photon pair sources that are utilized in these protocols are not suitable for large scale integration.\\

In general, the correlation enhanced target detection protocols (including QTMS) are inferior to the optimal entanglement-based protocols in terms of absolute performance. However, the correlation enhanced protocol only entails independent measurements instead of joint  measurement of the probe and reference photons, and is, therefore, not suitable for radar and ranging applications where the target distance is unknown.\\

In this paper, we have made a comparison with practically relevant CTI detection protocol that is in the optical domain, and not the optimal classical protocol. We believe it is still a fair comparison in the sense that the QTC detection protocol is also a practically realizable protocol. The proposed QTC detection protocol can be viewed as a separate practically realizable sensing technology made possible due to advances in single-photon detector technologies.  This new technology needs to be compared with current practical sensing technologies, particularly in the microwave. The theoretical results in this paper enable us to carry out such an analysis, which we plan to carry out in the future. 

\subsection{Practical applications}
A QTC source based on the proposed semiconductor waveguide approach is compact and practical. For instance, a moderate-size array could have the form factor of a laser pointer and be battery operable. Unlike many quantum technologies, no cryogenics is required. The accompanying detectors would be much bulkier and with significant power requirements, as such detectors need to be cooled. Application space could include covert ranging as well as covert imaging. However, significant signal processing work is required to fully exploit this capability, and the theory in this paper can be viewed as the first step of such an endeavor.

\section{Conclusion and Future Work}
We theoretically analyzed and experimentally demonstrated a prototype target detection protocol (the QTC detection protocol) with SPDC photon-pair sources.  This protocol is similar to the previously reported QTMS radar, but it utilizes the temporal correlation instead of the complex amplitude correlation between the probe and the reference photon and is in the optical regime (not microwave).
The QTC detection protocol only requires time-resolved photon-counting detection, which is phase-insensitive and therefore suitable for optical target detection.
As a comparison to the QTC detection protocol, we also consider a classical phase-insensitive target detection protocol based on intensity detection which is the basis of current technologies in the optical regime. \\

The experimental and theoretical results of this paper is summarized as follows:
\begin{itemize}
	\item The performance of both the CTI (and practical in the optical regime) and the QTC detection protocol are quantified with the standard ROC analysis that is widely adopted in radar/signal processing research. We present simple models and derive analytical expressions for the ROC curves, and demonstrate excellent agreement with experimental data. In particular, we applied the parameter estimation theory to construct the detection functions for target detection.\\

	\item We considered two different detector functions to quantify the performance of the QTC detection protocol. The QTC detection protocol is compared to the CTI detection protocol with intensity (photon counting) detection that is practically relevant in the optical regime. Experimental results show that in a lossy and noisy environment, the QTC detection protocol could achieve a probability of detection ($P_d$) of $\simeq 0.8$ for a false alarm probability around \(10^{-6}\), while the current classical/practical CTI detection protocol yields a $P_d\approx0$.\\

	\item This performance of the QTC detection protocol is comparable to the CTI detection protocol with the same output power but with detection time that is \(\simeq\)57 times longer.\\

\item The QTC detection protocol is also capable of covert ranging, with a low flux CW signal that is completely indistinguishable from the background noise. \\
\item We implemented a free-space target detection system with a semiconductor waveguide photon pair source. This demonstration provides an already proven, technological route to realizing a practical sensor in the optical regime.\\
\end{itemize}

There are many directions for further development of the QTC detection protocol. It would be important to model the probe photon transmission in greater detail for a specific QTC detection protocol, including the effect of target object reflectivity, range to target, the design of collection optics and the turbulence of the target detection channel. It is also important to explore the possibilities of array variants of the QTC detection protocol, i.e., with multiple sources and detectors to obtain a stronger probe photon flux. It is also important to investigate the possibility of overcoming the detector time uncertainty limit with novel photon detection techniques. These areas will be explored in future publications.
\bibliographystyle{IEEEtran}
\bibliography{./References_new}

\begin{thebibliography}{10}
\providecommand{\url}[1]{#1}
\csname url@samestyle\endcsname
\providecommand{\newblock}{\relax}
\providecommand{\bibinfo}[2]{#2}
\providecommand{\BIBentrySTDinterwordspacing}{\spaceskip=0pt\relax}
\providecommand{\BIBentryALTinterwordstretchfactor}{4}
\providecommand{\BIBentryALTinterwordspacing}{\spaceskip=\fontdimen2\font plus
\BIBentryALTinterwordstretchfactor\fontdimen3\font minus
  \fontdimen4\font\relax}
\providecommand{\BIBforeignlanguage}[2]{{%
\expandafter\ifx\csname l@#1\endcsname\relax
\typeout{** WARNING: IEEEtran.bst: No hyphenation pattern has been}%
\typeout{** loaded for the language `#1'. Using the pattern for}%
\typeout{** the default language instead.}%
\else
\language=\csname l@#1\endcsname
\fi
#2}}
\providecommand{\BIBdecl}{\relax}
\BIBdecl

\bibitem{greiner2013field}
W.~Greiner and J.~Reinhardt, \emph{Field quantization}.\hskip 1em plus 0.5em
  minus 0.4em\relax Springer Science \& Business Media, 2013.

\bibitem{einstein1935can}
A.~Einstein, B.~Podolsky, and N.~Rosen, ``Can quantum-mechanical description of
  physical reality be considered complete?'' \emph{Physical review}, vol.~47,
  no.~10, p. 777, 1935.

\bibitem{lloyd2008enhanced}
S.~Lloyd, ``Enhanced sensitivity of photodetection via quantum illumination,''
  \emph{Science}, vol. 321, no. 5895, pp. 1463--1465, 2008.

\bibitem{lopaeva:2013}
E.~Lopaeva, I.~R. Berchera, I.~Degiovanni, S.~Olivares, G.~Brida, and
  M.~Genovese, ``Experimental realization of quantum illumination,''
  \emph{Physical review letters}, vol. 110, no.~15, p. 153603, 2013.

\bibitem{BBDE2018}
B.~{Balaji} and D.~{England}, ``Quantum illumination: A laboratory
  investigation,'' in \emph{2018 International Carnahan Conference on Security
  Technology (ICCST)}, Oct 2018, pp. 1--4.

\bibitem{chang2019quantum}
C.~S. Chang, A.~Vadiraj, J.~Bourassa, B.~Balaji, and C.~Wilson,
  ``Quantum-enhanced noise radar,'' \emph{Applied Physics Letters}, vol. 114,
  no.~11, p. 112601, 2019.

\bibitem{Zhang:2015}
Z.~Zhang, S.~Mouradian, F.~N. Wong, and J.~H. Shapiro, ``Entanglement-enhanced
  sensing in a lossy and noisy environment,'' \emph{Physical review letters},
  vol. 114, no.~11, p. 110506, 2015.

\bibitem{Abolghasem:2009}
P.~Abolghasem, M.~Hendrych, X.~Shi, J.~P. Torres, and A.~S. Helmy, ``Bandwidth
  control of paired photons generated in monolithic bragg reflection
  waveguides,'' \emph{Optics letters}, vol.~34, no.~13, pp. 2000--2002, 2009.

\bibitem{caloz2018high}
M.~Caloz, M.~Perrenoud, C.~Autebert, B.~Korzh, M.~Weiss, C.~Sch{\"o}nenberger,
  R.~J. Warburton, H.~Zbinden, and F.~Bussi{\`e}res, ``High-detection
  efficiency and low-timing jitter with amorphous superconducting nanowire
  single-photon detectors,'' \emph{Applied Physics Letters}, vol. 112, no.~6,
  p. 061103, 2018.

\bibitem{klyshko1970parametric}
D.~Klyshko, A.~Penin, and B.~Polkovnikov, ``Parametric luminescence and light
  scattering by polaritons,'' \emph{Soviet Journal of Experimental and
  Theoretical Physics Letters}, vol.~11, p.~5, 1970.

\bibitem{zhukovsky2013analytical}
S.~V. Zhukovsky, L.~G. Helt, D.~Kang, P.~Abolghasem, A.~S. Helmy, and J.~Sipe,
  ``Analytical description of photonic waveguides with multilayer claddings:
  towards on-chip generation of entangled photons and bell states,''
  \emph{Optics Communications}, vol. 301, pp. 127--140, 2013.

\bibitem{liu2019enhancing}
H.~Liu, D.~Giovannini, H.~He, D.~England, B.~J. Sussman, B.~Balaji, and A.~S.
  Helmy, ``Enhancing lidar performance metrics using continuous-wave
  photon-pair sources,'' \emph{Optica}, vol.~6, no.~10, pp. 1349--1355, 2019.

\bibitem{ly2017tutorial}
A.~Ly, M.~Marsman, J.~Verhagen, R.~P. Grasman, and E.-J. Wagenmakers, ``A
  tutorial on fisher information,'' \emph{Journal of Mathematical Psychology},
  vol.~80, pp. 40--55, 2017.

\bibitem{marzban2004roc}
C.~Marzban, ``The {ROC} curve and the area under it as performance measures,''
  \emph{Weather and Forecasting}, vol.~19, no.~6, pp. 1106--1114, 2004.

\bibitem{QO}
\url{www.quantumopus.com}, the detail of the single photon detectors could be
  found at the Quantum Opus website. Accessed: 2019-08-08.

\bibitem{Horn:2012}
R.~Horn, P.~Abolghasem, B.~J. Bijlani, D.~Kang, A.~Helmy, and G.~Weihs,
  ``Monolithic source of photon pairs,'' \emph{Physical review letters}, vol.
  108, no.~15, p. 153605, 2012.

\bibitem{boyd2003nonlinear}
R.~W. Boyd, \emph{Nonlinear optics}.\hskip 1em plus 0.5em minus 0.4em\relax
  Elsevier, 2003.

\bibitem{vahlbruch2016detection}
H.~Vahlbruch, M.~Mehmet, K.~Danzmann, and R.~Schnabel, ``Detection of 15 db
  squeezed states of light and their application for the absolute calibration
  of photoelectric quantum efficiency,'' \emph{Physical review letters}, vol.
  117, no.~11, p. 110801, 2016.

\end{thebibliography}
\newpage
\begin{figure}[h!]
\includegraphics[width=1in,height=1.25in,clip,keepaspectratio]{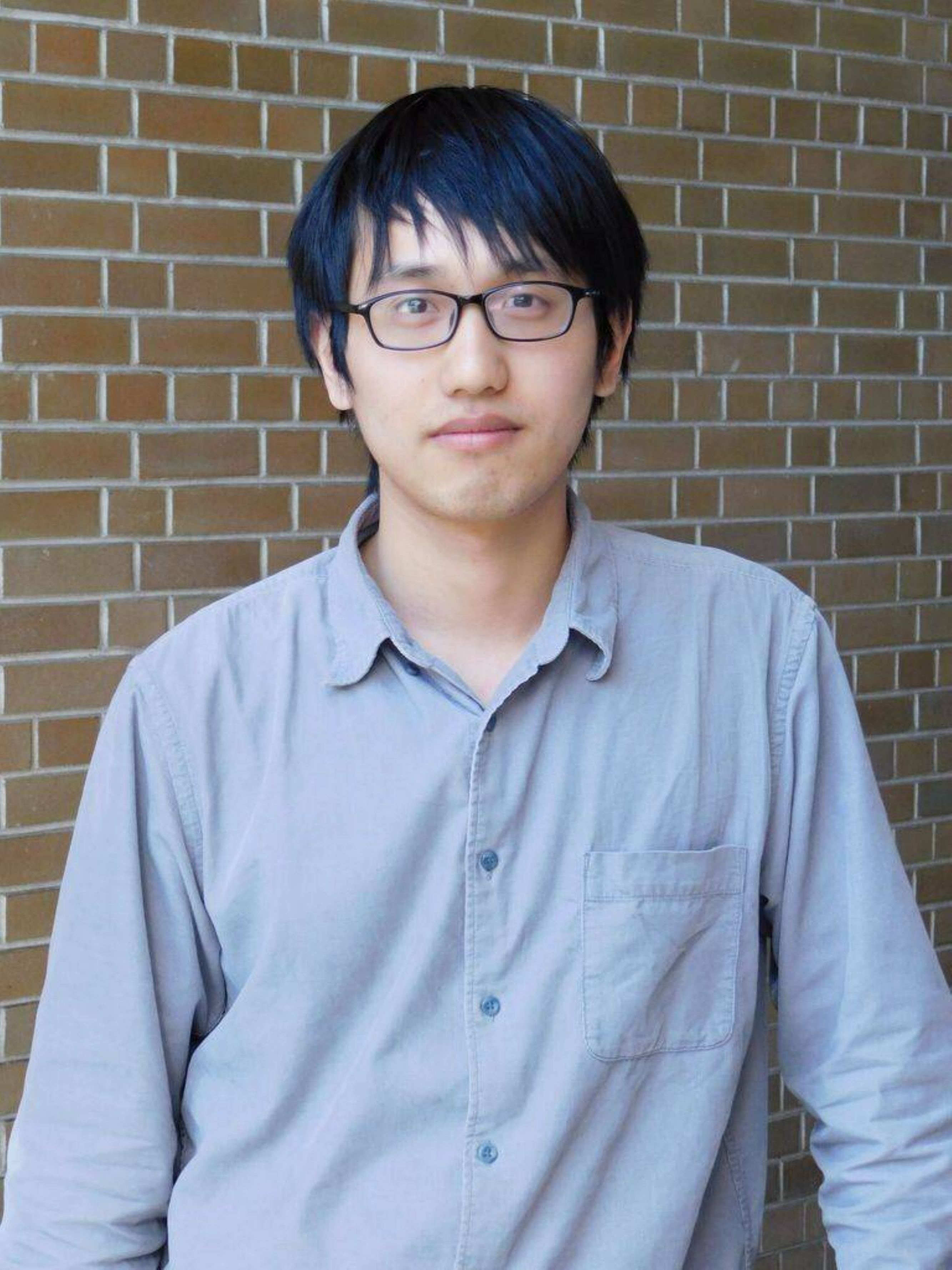}\\
\textbf{Liu Han} received his B.E. degree in opto-electronics information science and engineering from Tianjin university in 2016. He is now pursuing his Ph. D. degree in the university of Toronto. His research interests include quantum metrology and sensing with time-frequency entangled photon pairs.  
\end{figure}
\begin{figure}[h!]
\includegraphics[width=1in,height=1.25in,clip,keepaspectratio]{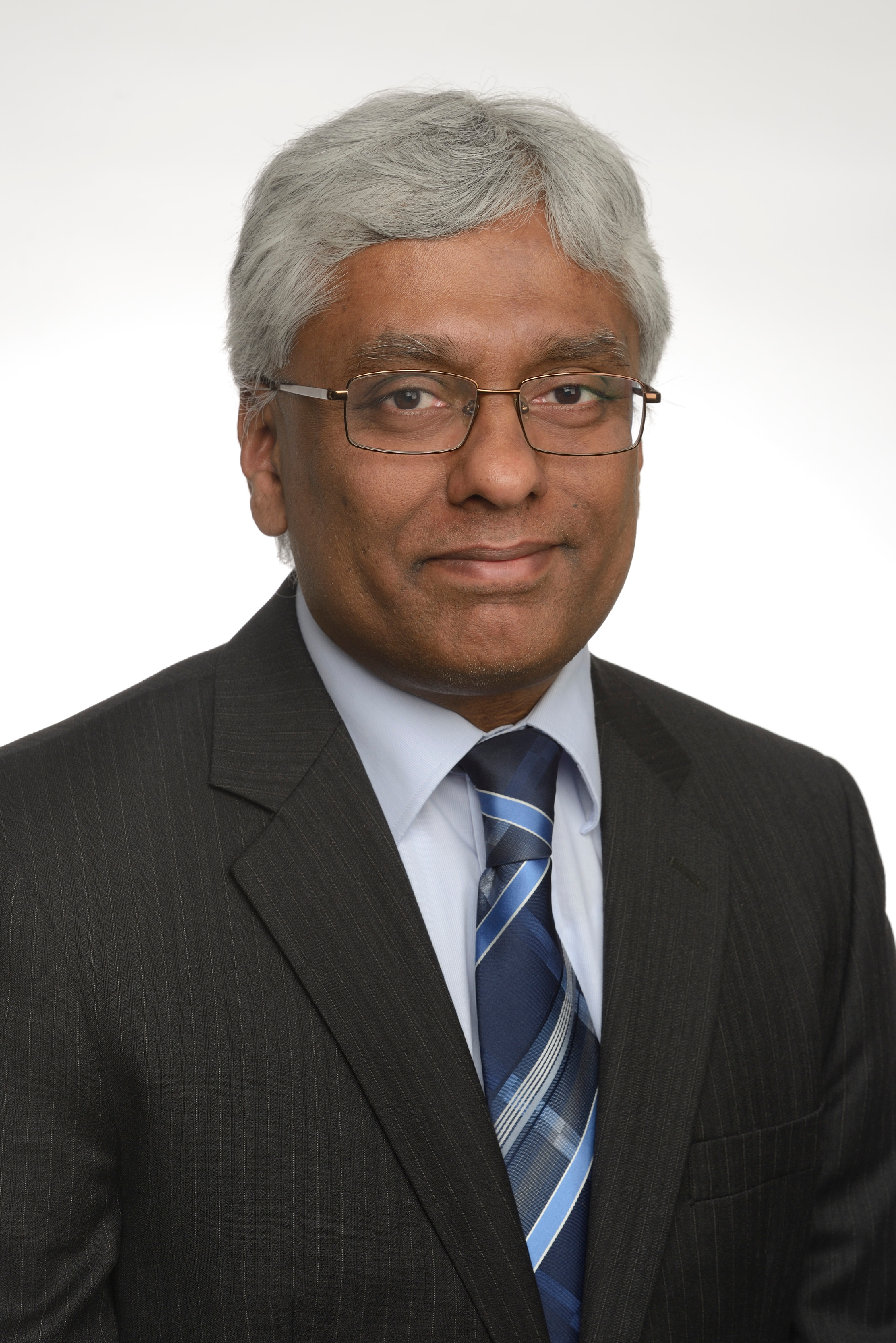}\\
\textbf{Bhashyam Balaji} (SMIEEE, FIET) graduated from Kendriya Vidyalaya (Central School), Sector VIII, Ram Krishna Puram, New Delhi, India, in 1987. He received his B.Sc. (Honors) degree in physics from St. Stephen’s College, University of Delhi, India, and his Ph.D. degree in theoretical particle physics from Boston University, Boston, MA, in 1997. Since 1998, he has been a scientist at the Defence Research and Development Canada, Ottawa, Canada. His research interests include all aspects of radar sensor outputs, including space-time adaptive processing, multi-target tracking, and meta-level tracking, as well as multisource data fusion. His theoretical research also includes the application of Feynman path integral and quantum field theory methods to the problems of nonlinear filtering and stochastic control. Most recently, his research interests include quantum sensing, in particular, quantum radar and quantum imaging. He is an IET Fellow and a recipient of the IEEE Canada Outstanding Engineer Award in 2018.
\end{figure}
\begin{figure}
\includegraphics[width=1in,height=1.25in,clip,keepaspectratio]{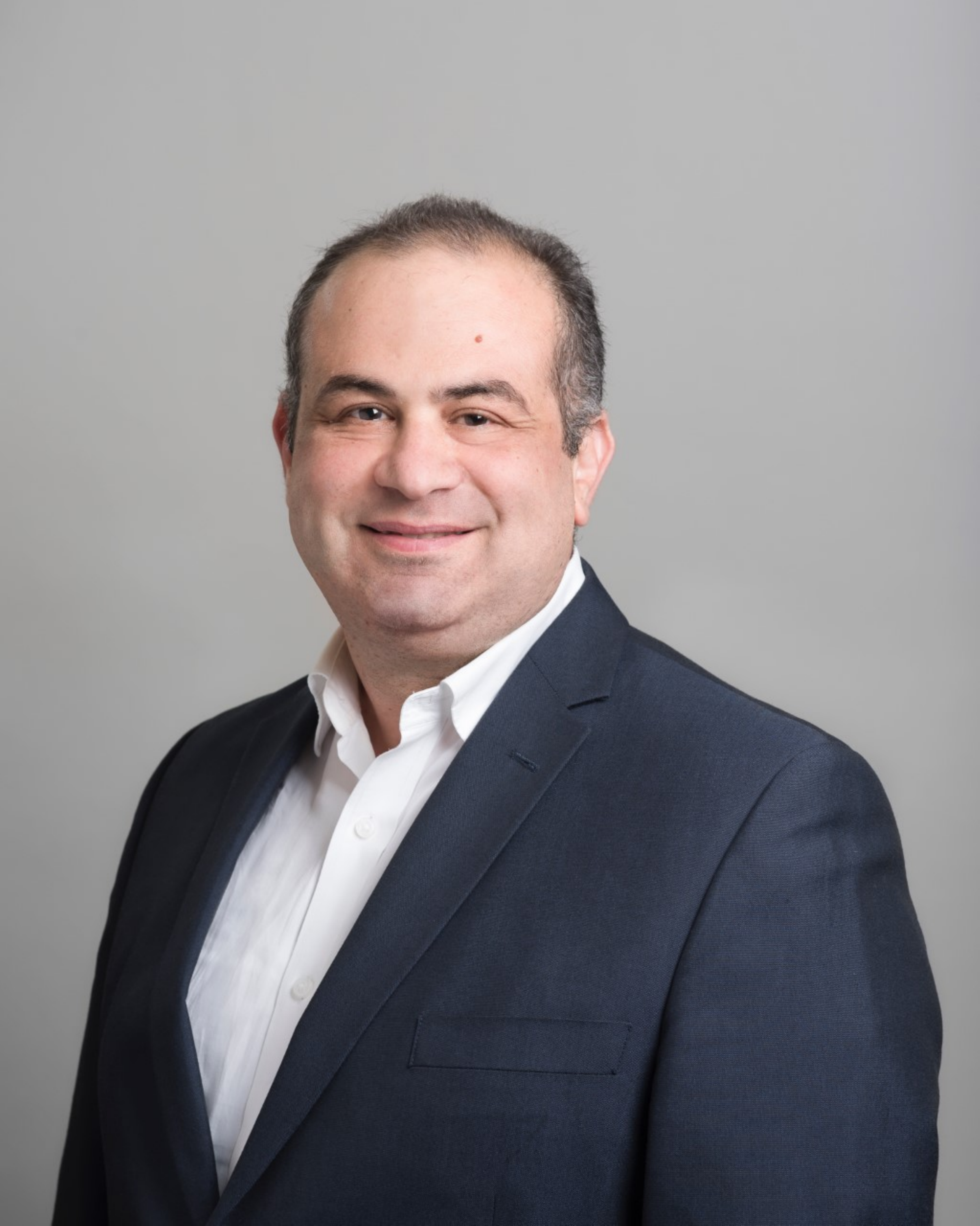}\\
	\textbf{Amr S. Helmy} is a Professor in the department of electrical and computer engineering at the University of Toronto.  Prior to his academic career, he held a position at Agilent Technologies photonic devices, R\&D division, in the UK between 2000 and 2004.  He received his Ph.D. and M.Sc. from the University of Glasgow with a focus on photonic devices and fabrication technologies, in 1999 and 1995 respectively.  He received his B.Sc. from Cairo University in 1993, in electronics and telecommunications engineering.\\

Amr is a senior member of the IEEE and a Fellow of the Optical Society of America. His research interests include photonic device physics and characterization techniques, with emphasis on nonlinear and quantum optics in III-V semiconductors; applied optical spectroscopy in III-V optoelectronic devices and materials; III-V fabrication and monolithic integration techniques.
\end{figure}
\newpage
\appendix
\subsection{Derivation of the photon counting statistics}
To calculate the coincidence detection rate \(P_c\), it suffices to consider a very small time interval \(\Delta \tau\) such that there could be at most one photon detection event on each detector. Then the probability of both the probe and the reference photons are generated and detected is given by \(\nu\eta_p\eta_r\Delta \tau\). In such cases, consider the temporal distribution of recorded probe and reference photon detection times $t_p$ and $t_r$ respectively. According to the definition of the detector time uncertainty \(\Delta t\), if the recorded detection time of the reference photon is given by \(t_r\), then the actual reference photon arriving time on the detector is within the range \([t_r-\frac{\Delta t}{2},t_r+\frac{\Delta t}{2}]\). And by the definition of the intrinsic correlation time \(\Delta t_0\), the actual probe photon arriving time on the probe detector is within the range of \([t_r-\frac{\Delta t}{2}-\Delta t_0,t_r+\frac{\Delta t}{2}+\Delta t_0]\). Again by the definition of detector time uncertainty, the recorded probe photon detection time is within the range \([t_r-\Delta t-\Delta t_0,t_r+\Delta t+\Delta t_0]\). Therefore the coincidence detection window width \(T_c\) must be set to twice the effective temporal uncertainty \(\Delta t_{eff} = \Delta t+\Delta t_0\) to ensure that every reference-probe photon pair that is detected will contribute to a coincidence detection event. Due to the non-zero width of the coincidence window \(T_c\), the noise photons and the reference photons will contribute to coincidence detection events too. The probability of detecting a reference photon and a noise photon that is within the coincidence window \(T_c\) is given by \(\nu\eta_r\nu_bT_c\Delta \tau\). The total rate of coincidence detection events is given by:
\begin{align}
    P_c &= \frac{\eta_p\eta_r\nu\Delta \tau+\eta_r\nu\nu_bT_c\Delta \tau}{\Delta \tau}\\
        &= \eta_p\eta_r\nu+\eta_r\nu\nu_bT_c
\end{align}
The derivation of single channel detection event rates \(P_p,P_r\) is directly from the definition:
\begin{align}
    P_p &= \eta_p\nu+\nu_b-P_c\\
    P_r &= \eta_r\nu-P_c
\end{align}

For a target detection experiment that last time \(\tau\), we make an important approximation that \(\mathbf{N}_p\), \(\mathbf{N}_r\) and \(\mathbf{N}_c\) each obey independent Poisson distribution, with mean value \(P_p\tau\), \(P_r\tau\) and \(P_c\tau\).
The independent Poisson distribution approximation is valid for the following reasons.
The single channel detection events and coincidence detection events are completely uncorrelated if the time separation is larger than the effective time uncertainty \(\Delta t_{eff}\) (assuming the optical path length difference between reference and probe photon is zero).
This is because the probe photon is only correlated to the reference photon from the same SPDC photon pair.
Secondly, the probability of having more than one single channel or coincidence detection events within an effective time uncertainty \(\Delta t_{eff}\) is negligible.
This is because the photon detection rate is much lower than the inverse of the effective time uncertainty.
Based on these two reasons, single-channel detection events and coincidence detection events could be treated as if completely uncorrelated and therefore each obeys independent Poisson distribution.\\

\subsection{Derivation of the Fisher information expressions}
For a discrete random variable \(\mathbf{X}\) whose probability mass function \(P(\mathbf{X}=x|\eta)\) is parametrized by a parameter \(\eta\), the Fisher information of \(\mathbf{X}\) about \(\eta\) is given by:
\begin{align}
I_{\mathbf{X}}(\eta) = \sum\limits_{x \in \chi} \left[\frac{d}{d\eta}\text{log}P(\mathbf{X}=x|\eta)\right]^2 P(\mathbf{X}=x|\eta)
\end{align}
where \(\chi\) is the space of different outcome of \(\mathbf{X}\) and for the QTC detection experiment \(\chi\) is given by:
\begin{gather}
\chi = \{(N_p,N_r,N_c)|N_p,N_r,N_c \in \mathbb{N}\}
\end{gather}
To simplify the derivation of the Fisher information for the QTC detection protocol, we show first the additive property of Fisher information \(I_{xy}\). Consider a joint probability distribution of two independent random variables \(f_{xy}(x,y) = f_x(x)f_y(y)\):
\begin{align}
I_{xy} &= \sum\limits_{x,y}\left[\frac{d}{d\eta}\text{log}f_{xy}(x,y)\right]^2f_{xy}(x,y)\\  \nonumber
&= \sum\limits_{x,y}\frac{f_x(x)f_y(y)}{[f_x(x)f_y(y)]^2}\left[\d{f_x(x)}{\eta}f_y(y)+\d{f_y(y)}{\eta}f_x(x)\right]\\ \nonumber
&= \sum\limits_{x,y}\frac{1}{f_x(x)f_y(y)}\left[\d{f_x(x)}{\eta}f_y(y)+\d{f_y(y)}{\eta}f_x(x)\right]^2\\ \nonumber
&= \sum\limits_{x,y}\{[\d{f_x(x)}{\eta}]^2\frac{f_y(y)}{f_x(x)}+[\d{f_y(y)}{\eta}]^2\frac{f_x(x)}{f_y(y)}\\ \nonumber
&\quad+2\d{f_x(x)}{\eta}\d{f_y(y)}{\eta}\}\\ \nonumber
&= \sum\limits_{x}\left\{\frac{1}{f_x(x)}\left[\d{f_x(x)}{\eta}\right]^2+\sum\limits_{y}\frac{1}{f_y(y)}\left[\d{f_y(y)}{\eta}\right]^2\right\}\label{TEMP}\\ \nonumber
&= \sum\limits_{x}\left[\frac{d}{d\eta}\text{log}f_x(x)\right]^2f_x(x)\\  \nonumber
&\quad+ \sum\limits_{y}\left[\frac{d}{d\eta}\text{log}f_y(y)\right]^2f_y(y)\\
&=I_x+I_y
\end{align}

In \eqref{TEMP} the normalization property of probability distribution (\(f_z(z),z\in \{x,y\}\)) is used:
\begin{align}
\sum\limits_z f_z(z) &= 1\\
\sum\limits_z \d{f_z(z)}{\eta}& = 0
\end{align}
Therefore, to calculate the Fisher information \(I_\text{QTC1}\) for the joint probability distribution \(P(\mathbf{N}_p,\mathbf{N}_r,\mathbf{N}_c)\), which is a product of independent Poisson distributions, it suffices to calculate the sum of Fisher information of the probability distributions of \(\mathbf{N}_p\), \(\mathbf{N}_r\) and \(\mathbf{N}_c\):
\begin{align} \label{FT_3}
I_\text{QTC1} &= \sum\limits_{N_p,N_r,N_c=0}^{+\infty}\\ \nonumber
    &\quad    P(\mathbf{N}_p=N_p,\mathbf{N}_r=N_r,\mathbf{N}_c=N_c)\\ \nonumber
&\quad\times\left[\pd{}{\eta_p}\text{log}P(\mathbf{N}_p=N_p,\mathbf{N}_r=N_r,\mathbf{N}_c=N_c\right]^2\\ \nonumber
&=\sum\limits_{N_p=0}^{+\infty} \left[\pd{}{\eta_p}\text{log}f(N_p,P_p\tau)\right]^2f(N_p,P_p\tau)\label{FT1}\\ \nonumber
&\quad+\sum\limits_{N_r=0}^{+\infty} \left[\pd{}{\eta_p}\text{log}f(N_r,P_r\tau)\right]^2f(N_r,P_r\tau)\\ \nonumber%\label{FT2}\\ \nonumber
&\quad+\sum\limits_{N_c=0}^{+\infty} \left[\pd{}{\eta_p}\text{log}f(N_c,P_c\tau)\right]^2f(N_c,P_c\tau)
\end{align}
It can be shown that for Poisson distribution
\begin{align}
f(k,\lambda) = e^{-\lambda}\frac{\lambda^k}{k!}
\end{align}
whose mean value \(\lambda\) is parametrized by \(\eta\), the Fisher information is given by:
\begin{align}
 I_k &= \sum\limits_{k=0}^{+\infty}\left[\pd{}{\eta_p}\text{log}f(k,\lambda)\right]^2f(k,\lambda)\\ \nonumber
 &=\sum\limits_{k=0}^{+\infty}\frac{1}{f(k,\lambda)}\left[\d{f(k,\lambda)}{\lambda}\d{\lambda}{\eta}\right]^2\\ \nonumber
 &=\frac{1}{\lambda^2}\left[\d{\lambda}{\eta}\right]^2\sum\limits_{k=0}^{+\infty}e^{-\lambda}\frac{\lambda^k}{k!}(k-\lambda)\\ \nonumber
 &=\frac{1}{\lambda}\left[\d{\lambda}{\eta}\right]^2\label{FP} \end{align}
Then from \eqref{FT1} we obtain%-\eqref{FT3}
\begin{align}
I_\text{QTC1} &= \frac{1}{P_p\tau}\left[\d{P_p}{\eta_p}\tau\right]^2+\frac{1}{P_r\tau}\left[\d{P_r}{\eta_p}\tau\right]^2+\frac{1}{P_c\tau}\left[\d{P_c}{\eta_p}\tau\right]^2,\\ \nonumber &=\tau\nu^2\left[\frac{(1-\eta_r)^2}{P_p}+\frac{\eta_r^2}{P_r}+\frac{\eta_r^2}{P_c}\right]
\end{align}

The Fisher information \(I_\text{QTC2}\) and \(I_\text{CTI}\) is obtained from Poisson distribution \(P(\mathbf{N}_c=N_c) = f(N_c,P_c\tau)\) and \(P(\mathbf{N}_p=N_p) = f(N_p,P_p\tau)\). Then it directly follows from \eqref{FP} that:
\begin{align}
I_\text{QTC2} &= \frac{\eta_r^2}{P_c}\tau\nu^2
\end{align}
and
\begin{align}
 I_\text{CTI} &= \frac{(1-\eta_r)^2}{P_p}\tau\nu^2\Big|_{\eta_r=0}\\  \nonumber
 &=\frac{\nu^2}{P_p}\tau.
 \end{align}
\newpage
 \section*{figure captions}
\textbf{Fig. \ref{JTI}} Top: the schematic of the SPDC process in the semiconductor waveguide. The red wave \(\lambda_0\) is the pump light at 783nm, the green \(\lambda_p\) and blue \(\lambda_r\) wave is the down-converted  probe and reference photon that is in horizontal (vertical) polarization. Middle: joint probability distribution of photon detection time.
\(t_p\): detection time of the probe photon; \(t_r\) detection time of the reference photon.
The standard deviation of the detection time difference is \(\Delta (t_p-t_s) = 100ps\) for this plot. Bottom: Block diagram of the experiment setup.\\

\textbf{Fig. \ref{SETUP}} The schematic of the experimental setup is divided into two parts: the source and detection part and the transceiver part. The source and detection part (green background) includes the photon pair source and detectors. Pump laser: Ti-Sapphire CW laser at 783nm. PBS: polarization beam splitter. LPF: long pass (\(>\)1200 nm) filter to separate the SPDC photon pairs from the pump laser. SNSPD: dual-channel superconducting single-photon detector (top channel: the reference detector, bottom channel: the probe detector). Transceiver part (blue background): probing of the target object and collection of the back-reflected photons. Target object: a piece of aluminum foil with diffused reflection. The source and detector part and the transceiver part are built on a separate table and are connected by single-mode fibers (yellow line with arrow).\\

\textbf{Fig. \ref{PIC}} Photograph of the transceiver part of the experiment setup. Different optical elements are marked with red boxes. The green line marks the optical path of the probe photon before it hits on the target object. The reflective mirror on the bottom-right corner is purely for optical alignment purposes.\\

\textbf{Fig. \ref{SEM}}the scanning electron microscope image of the semiconductor waveguide\\

\textbf{Fig. \ref{TS_HIST}}Left:the time series of different detector functions \(\mathbf{S}_\text{QTC1}\), \(\mathbf{S}_\text{QTC2}\) and \(\mathbf{S}_\text{CTI}\) over 1400 independent target experiment.
X-axis: the index of independent experiment that last time \(\tau=0.01\)s.
Blue (red) bar: probability density when the target is absent (present).
Blue (red) solid line, the theoretically calculated probability distribution using the Gaussian signal approximation \eqref{GAPPROX}. The histograms are calculated from a larger portion of data (\(2.19\times10^{6}\) independent experiments).\\

\textbf{Fig. \ref{ROC_FIG}} Left:ROC curves for the detector function \(\mathbf{S}_\text{QTC1}\), \(\mathbf{S}_\text{QTC2}\) and \(\mathbf{S}_\text{CTI}\) for \(\tau=0.02\)s, \(\tau=0.01\)s and \(\tau_=0.005\)s (from top to bottom).
Red curve: the theoretical ROC curve for \(\mathbf{S}_\text{QTC1}\) calculated based on the Gaussian signal approximation.
Red triangles:the experimental ROC curve for \(\mathbf{S}_\text{QTC1}\).
The blue curve, blue triangles, and black curve, black triangles are similarly plotted for detector functions \(\mathbf{S}_\text{QTC2}\) and \(\mathbf{S}_\text{CTI}\), respectively.
Dashed black curve: theoretical ROC curve for the detector function \(\mathbf{S}_\text{CTI}\) with longer integration time \(\tau'=k_\tau\tau\) such that the detection rate \(P_d\) same as that of \(\mathbf{S}_\text{QTC1}\) is achieved at false alarm rate \(P_{fa} =10^{-6}\). The ratios are \(k_\tau=42.62,56.75,80.55\) for \(\tau=0.005s,0.01s,0.02s\). For the same integration time \(\tau\), the CTI detection protocol can also achieve the same performance with \(6.59,7.59,9.01\) times more source power or total probe photon transmission.
Right: the same ROC curve plotted with log scaled \(P_d\) axis.\\

\textbf{Fig. \ref{ROC_FIG_OPT}} Theoretical ROC curves with measurement time \(\tau=0.005\)s and improved effective time uncertainty \(\Delta t_{eff}\) or reference photon transmission \(\eta_r\) for the QTC detection protocol. Black: the CTI detection protocol for comparison; Red: the QTC detection protocol (\(\mathbf{S}_\text{QTC1}\)) without any improvement; Black dashed: the CTI detection protocol with \(\simeq80.55\) times longer measurement time; Red dashed: the QTC detection protocol with improved effective time uncertainty \(\Delta t_{eff}=25 ps\); Red dotted: the QTC detection protocol with perfect reference photon transmission \(\eta_r=100\%\).

\end{document}